\documentclass[aps,prb,a4paper,showpacs,showkeys,twocolumn]{revtex4-1}
\usepackage{graphicx}
\usepackage{amsmath}
\usepackage{dcolumn}
\usepackage{subfig}

\newcommand{\me}{\text{e}}
\newcommand{\dif}{\mathrm{d}}

\begin{document}
\title{Material configurations for $n$-type silicon-based terahertz quantum 
cascade lasers}
\author{A. Valavanis}
\email{a.valavanis@leeds.ac.uk}
\author{T. V. Dinh}
\author{L. J. M. Lever}
\author{Z. Ikoni\'{c}}
\author{R. W. Kelsall}
\affiliation{Institute of Microwaves and Photonics, School of Electronic
and Electrical Engineering, University of Leeds, Leeds LS2 9JT, United
Kingdom}
\date{\today}

\begin{abstract}
Silicon-based quantum cascade lasers (QCLs) offer the prospect of integrating 
coherent THz radiation sources with silicon microelectronics. Theoretical
studies have proposed a variety of $n$-type SiGe-based heterostructures as
design candidates, however the optimal material configuration remains unclear.
In this work, an optimization algorithm is used to design equivalent THz QCLs in 
three recently-proposed configurations [(001) Ge/GeSi, (001) Si/SiGe 
and (111) Si/SiGe], with emission frequencies of 3 and 4\,THz.
A systematic comparison of the electronic and optical properties is presented. A
semi-classical electron transport simulation is used to model the charge carrier
dynamics and calculate the peak gain, the corresponding current density and the
maximum operating temperature. It is shown that (001) Ge/GeSi structures yield
the best simulated performance at both emission frequencies.
\end{abstract}



\maketitle

\section{Introduction}
Terahertz quantum cascade lasers (THz QCLs) are semiconductor devices 
in which electrons are transported through a periodic multiple quantum well 
heterostructure, with a radiative transition in each period.\cite{NatureKohler2002}
THz QCLs have numerous potential applications including radiation sources for
medical and security imaging, and local oscillators in astronomy and remote gas 
sensing.\cite{NatPhotonicsWilliams2007, NatureHolland1998,
OptExpLoffler2001, PRBCalifano2007, NatureGraham2007}

All THz QCLs to date have been fabricated from III--V compound 
semiconductors (for example Ref.~\onlinecite{APLKumar2009, APLAjili2005}).
However, Si-based QCLs could
offer a number of significant advantages.  Mature Si
processing technology may reduce costs and allow integration with conventional
electronic devices.  Existing THz QCLs operate only at cryogenic temperatures
(currently below 186\,K for resonant-phonon QCLs\cite{APLKumar2009} or 116\,K
for bound--to--continuum\cite{APLScalari2005}), but the high thermal conductivity 
of Si-based structures could enable heat to be extracted more effectively, and 
hence allow higher
operating temperatures.  III--V QCLs are also limited to THz emission
at frequencies lower than 4.9\,THz,\cite{APLLee2006} owing to the strong
absorption in the Reststrahlen band.  However, this limitation does not exist in 
non-polar group-IV materials.

Although the indirect bandgap in Si has, so far, frustrated efforts to develop 
an interband laser, this is not an issue for intersubband devices such as 
QCLs because the radiative transitions occur between subbands within the same
valley of an energy band.
Mid-infrared\cite{ScienceDehlinger2000} and THz\cite{APLLynch2002} intersubband
electroluminescence has been observed from $p$-type SiGe/Si quantum cascade 
structures.  However, dispersion relations for holes in these structures are
quite complicated, owing to the contributions from multiple valence 
bands, and in recent years, attention has switched toward $n$-type structures.  
This greatly simplifies the device design process and may result in a lower
spectral linewidth (and hence greater peak gain) than that of $p$-type
structures. Early design proposals included $n$-type Si/SiGe structures that
exploited transitions in the $\Delta$ valleys of (001) Si quantum wells (QWs), 
with SiGe barriers.\cite{IJHSESKelsall2003} 
However, a range of alternative material configurations have been considered
in recent years.  $L$ valley transitions in
(001) oriented Ge/GeSi have attracted greatest attention,\cite{APLLever2009,
APLDriscoll2006, JAPDriscoll2007} although transitions in the $\Delta$ valleys
of (111) oriented Si/SiGe,\cite{PRBValavanis2008_2, APLLever2008} the
$\Gamma$ valley of Ge/GeSi,\cite{SSTHan2007} and the $L$ valleys of
Ge/GeSiSn\cite{APLSun2007} have also been considered.

Many properties relating to the bandstructure
and carrier transport have a strong effect upon the gain of QCLs. To date, no
quantitative comparison has been made of the expected performance of THz QCLs in
different Si-based material configurations. In this paper, we present such a
comparison by simulating equivalent devices in the (001) Si/SiGe, (111) Si/SiGe
and (001) Ge/GeSi material configurations, using a detailed semi-classical
rate-equation approach.  Devices emitting near 3 and 4\,THz were designed for
each material configuration by rescaling a recent bound--to--continuum 
design\cite{APLLever2009} according to the effective mass of the material and 
then applying an automated design-optimization algorithm.
In section~\ref{scn:bandstructure}, we calculate the conduction band potentials
for each material configuration and the range of energies within QWs that can be
used for QCL design.  
Section~\ref{scn:transport} describes our model for simulating carrier 
transport, gain, and current density in QCLs.  The design-optimization algorithm
is described in section~\ref{scn:devices} and a summary of the optimized 
devices is presented.  Finally, section~\ref{scn:results} presents a comparison
of the simulated performance of devices in each material system.

\section{Conduction band states}
\label{scn:bandstructure}
\begin{table}[t]
  \caption{\label{tbl:MaterialConstants}Material parameters for Si and Ge.}
    \begin{ruledtabular}
    \begin{tabular}{lddl}
      Constant & \text{Si} & \text{Ge} & Unit\\
      \hline
      \hline
      $a$ 
          & 0.5431\footnotemark[1]
          & 0.5633\footnotemark[1]
          & nm\\
      $\Delta_{\mathrm{so}}$
          & 44.0\footnotemark[2]
          & 296.0\footnotemark[3]
          & meV\\
      $C_{11}$
          & 165.773\footnotemark[4]
          & 128.528\footnotemark[4]
          & GPa\\
      $C_{12}$
          & 63.924\footnotemark[4]
          & 48.260\footnotemark[4]
          & GPa\\
      $C_{44}$ 
          & 79.619\footnotemark[4]
          & 66.799\footnotemark[4]
          & GPa\\
      $(\Xi_d + \frac{1}{3}\Xi_u - a_v)^{\Delta}$
          & 1.72\footnotemark[5]
          & 1.31\footnotemark[5]
          & eV\\
      $(\Xi_d + \frac{1}{3}\Xi_u - a_v)^{L}$
          & -3.12\footnotemark[5]
          & -2.78\footnotemark[5]
          & eV\\
      $\Xi_{u}^{\Delta}$
          & 9.16\footnotemark[5]
          & 9.42\footnotemark[5]
          & eV\\
      $\Xi_{u}^{L}$
          & 16.14\footnotemark[5]
          & 15.13\footnotemark[5]
          & eV
      \footnotetext[1]{Reference~\onlinecite{JPhysChemDismukes1964}}
      \footnotetext[2]{Reference~\onlinecite{PRLZwerdling1960}}
      \footnotetext[3]{Reference~\onlinecite{JPCSKane1956}}
      \footnotetext[4]{Reference~\onlinecite{JAPMcSkimin1964}}
      \footnotetext[5]{Reference~\onlinecite{PRBVanDeWalle1986}}
      \end{tabular}
    \end{ruledtabular}
\end{table}

\subsection{Model solid approximation}

The model solid approximation\cite{PRBVanDeWalle1986} was used to calculate the
conduction band offset between a strained Si$_{1-x}$Ge$_x$ alloy and a 
substrate material.  The difference between the average of the 
light-hole, heavy-hole, and spin-orbit split off valence band edges in the two
materials was used as a reference energy, as it is almost independent of strain 
and crystal orientation.\cite{PRBVanDeWalle1986} 
The value of this property was interpolated
from empirical pseudopotential data as\cite{PRBRieger1993}
\begin{equation}
  \Delta\overline{E_v} = (0.47 - 0.06x_s)(x-x_s),
\end{equation}
where $x_s$ is the Ge fraction in the substrate.  
The valence band maximum in an unstrained bulk alloy was found using
\begin{equation}
  E_v^{\text{bulk}} = \Delta\overline{E_v} + \frac{1}{3}\Delta_{\text{SO}},
\end{equation}
where $\Delta_{\text{SO}}$ is the spin-orbit splitting energy.  Material 
parameters for alloys were found by interpolating from the bulk Si and Ge 
values in table~\ref{tbl:MaterialConstants}.
These parameters yield an offset of 0.55\,eV between the top of the valence
bands in bulk Ge and Si, which matches a recently-measured value for 
weakly-strained Ge and Si films.\cite{APLAfanas'ev2009}  Very similar parameters
have also been shown to yield close agreement with experimental measurements of 
intersubband absorption energies in Ge/GeSi QWs.\cite{PRBBusby2010}

The low-temperature indirect bandgaps for the $\Delta$ and $L$ valleys
in an unstrained alloy (in eV) were taken as\cite{PRBWeber1989}
\begin{eqnarray}
  E_g^{\Delta} &=& 1.155 - 0.43x + 0.0206x^2\\\nonumber
  E_g^{L}      &=& 2.010 - 1.270x,
\end{eqnarray}
and the unstrained conduction band edge for a given valley was found using
$E_c^{\text{bulk}} = E_v^{\text{bulk}} + E_g$.

The effects of hydrostatic and uniaxial strain on the band edge were determined
as follows.
First, the lattice constant of a thin Si$_{1-x}$Ge$_x$ layer was found using
\begin{equation}
  a(x)=a_\mathrm{Si}(1-x)+a_\mathrm{Ge}x-b_\mathrm{bow}x(1-x),
\end{equation}
where $b_\mathrm{bow}=0.2733$\,pm\cite{JPhysChemDismukes1964} is a bowing 
factor.\cite{SSTPaul2004}  The layer was assumed to deform uniformly, such that 
the lattice constant matched that of a thick substrate material, $a_s$.
The resulting strain in the plane of epitaxial growth was given
by $\varepsilon_{\parallel} = (a_s-a)/a$.

The hydrostatic deformation was found for each set of conduction band
valleys using
\begin{equation}
    \Delta E_g^{\text{Hyd}} 
        = \left(\Xi_d + \frac{1}{3}\Xi_u 
        - a_v\right)\operatorname{Tr}{\mathsf{\varepsilon}'},
\end{equation}
where $\Xi_d + \frac{1}{3}\Xi_u - a_v$ is the bandgap deformation potential for
the $\Delta$ or $L$ valleys and $\operatorname{Tr}{\mathsf{\varepsilon}'}$ is
the trace of the strain tensor, where\cite{SSESmirnov2004}
\begin{eqnarray}
    \operatorname{Tr}{\mathsf{\varepsilon}'^{(001)}} &=& 2\left(1-\frac{C_{12}}{C_{11}}\right)\varepsilon_{\parallel}\\
    \operatorname{Tr}{\mathsf{\varepsilon}'^{(111)}} &=& \frac{12C_{44}}{C_{11}+2C_{12}+4C_{44}}\varepsilon_{\parallel}
\end{eqnarray}
for (001) and (111) oriented epilayers, respectively.  In the above equations,
$C_{11}, C_{12}$, and $C_{44}$ are elastic constants.

Uniaxial strain leads to splitting of the $\Delta$ valley degeneracy in 
(001) oriented layers.  The energy shifts are given by\cite{SSESmirnov2004}
\begin{eqnarray}
  \Delta E_{c}^{\Delta_4,\text{Uni}}
      &=& \frac{1}{3}\Xi_u^{\Delta}\left(1 + \frac{2C_{12}}{C_{11}}\right)\varepsilon_{\parallel} \\\nonumber
  \Delta E_{c}^{\Delta_2,\text{Uni}}
      &=& -\frac{2}{3}\Xi_u^{\Delta}\left(1 + \frac{2C_{12}}{C_{11}}\right)\varepsilon_{\parallel},
\end{eqnarray}
for the valleys with their major axes perpendicular and parallel to the 
growth-direction respectively, where $\Xi_u^{\Delta}$ is the uniaxial
deformation potential for the $\Delta$ valleys. Similarly, $L$ valleys in (111)
oriented layers are shifted by
\begin{eqnarray}
  \Delta E_{c}^{L_1,\text{Uni}}
      &=& -2\Xi_u^L \frac{C_{11} + 2C_{12}}{C_{11} + 2C_{12} + 4C_{44}}   \varepsilon_{\parallel}\\\nonumber
  \Delta E_{c}^{L_3,\text{Uni}}
      &=& \frac{2}{3}\Xi_u^L\frac{C_{11} + 2C_{12}}{C_{11} + 2C_{12} + 4C_{44}}   \varepsilon_{\parallel},
\end{eqnarray}
for the valley with its major axis in the growth direction, and the three other 
valleys respectively.  Uniaxial strain has no effect upon the $\Delta$ valleys
in (111) layers, or the $L$ valleys in (001) layers, owing to symmetry.

Finally, the energy of a given conduction band minimum in a strained
layer was found relative to the average substrate valence band using
\begin{equation}
  E_c = E_c^{\text{bulk}} + \Delta E_g^{\text{Hyd}} 
      + \Delta E_c^{\text{Uni}}.
\end{equation}

\subsection{Available energy range}
\begin{figure}
\includegraphics*[width=8.6cm]{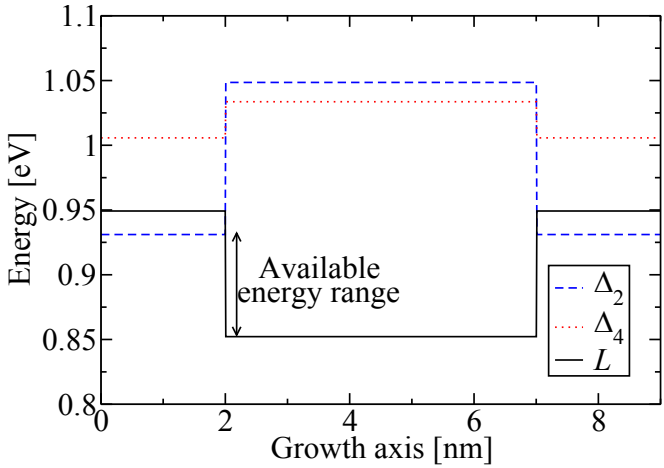}
\caption{(Color online) Spatial variation of $\Delta$ and $L$ conduction band 
minima in a Si$_{0.15}$Ge$_{0.85}$/Ge/Si$_{0.15}$Ge$_{0.85}$ heterostructure on 
a Si$_{0.025}$Ge$_{0.975}$ virtual substrate.  The energy range available for
QCL design is defined here as the region between the bottom of the QW and the
next-lowest conduction band valley.}
\label{fig:WellDepth}
\end{figure}

In this section, we determine the energy ranges within group-IV QWs that can 
be reliably exploited for THz QCL design. It is insufficient to simply calculate
the depth of a QW as there are multiple conduction band valleys within the
energy range of interest.  This can degrade device performance by introducing 
undesirable intervalley scattering processes.  To avoid this problem, we 
consider the energy difference between the bottom of the well,
and the next-lowest conduction band minimum (which may be in either the well or
the barrier). This is illustrated in Fig.~\ref{fig:WellDepth} for the case of a
Si$_{0.15}$Ge$_{0.85}$/Ge/Si$_{0.15}$Ge$_{0.85}$ QW on a
Si$_{0.025}$Ge$_{0.975}$ substrate. Here, the $L$ valleys form the bottom of the
well and the usable energy range is limited by the $\Delta_2$ valley minima in
the barriers.

\begin{figure*}
	\subfloat[][]{
		\label{subfig:usable-energy-Si001}
		\includegraphics*[width=0.3\textwidth]{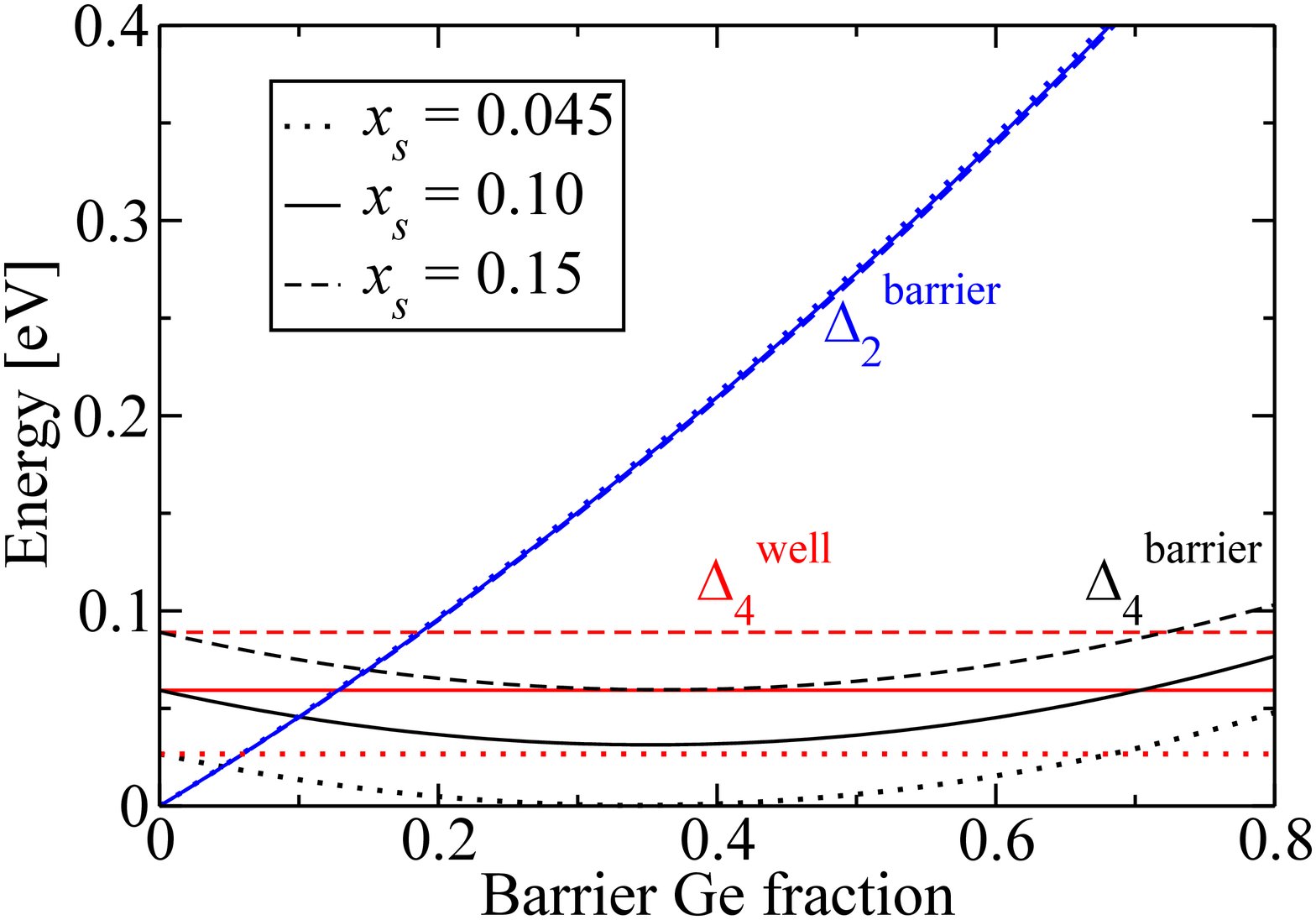}
		}
	\subfloat[][]{
		\label{subfig:usable-energy-Si111}
		\includegraphics*[width=0.3\textwidth]{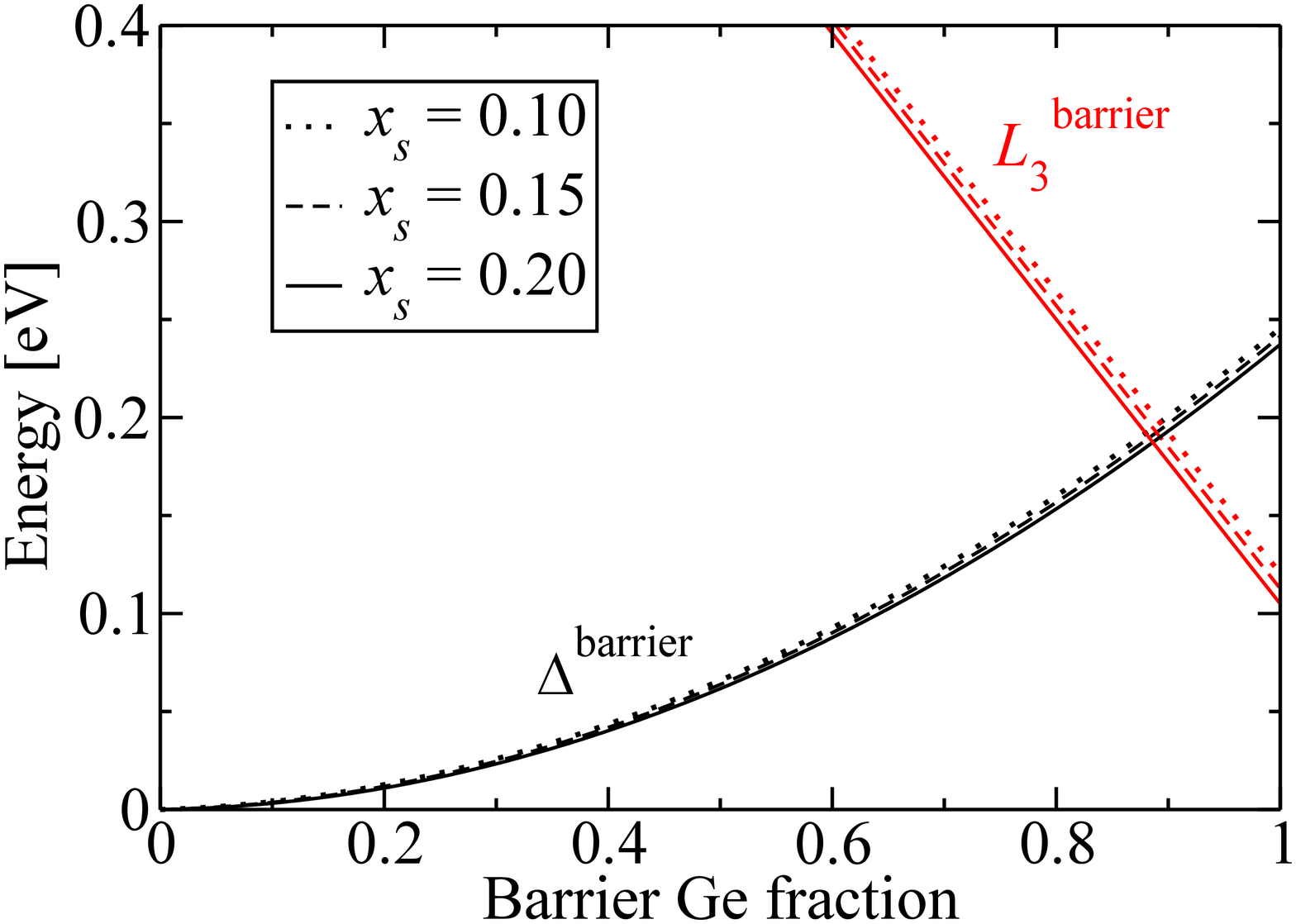}
		}
	\subfloat[][]{
		\label{subfig:usable-energy-Ge001}
		\includegraphics*[width=0.3\textwidth]{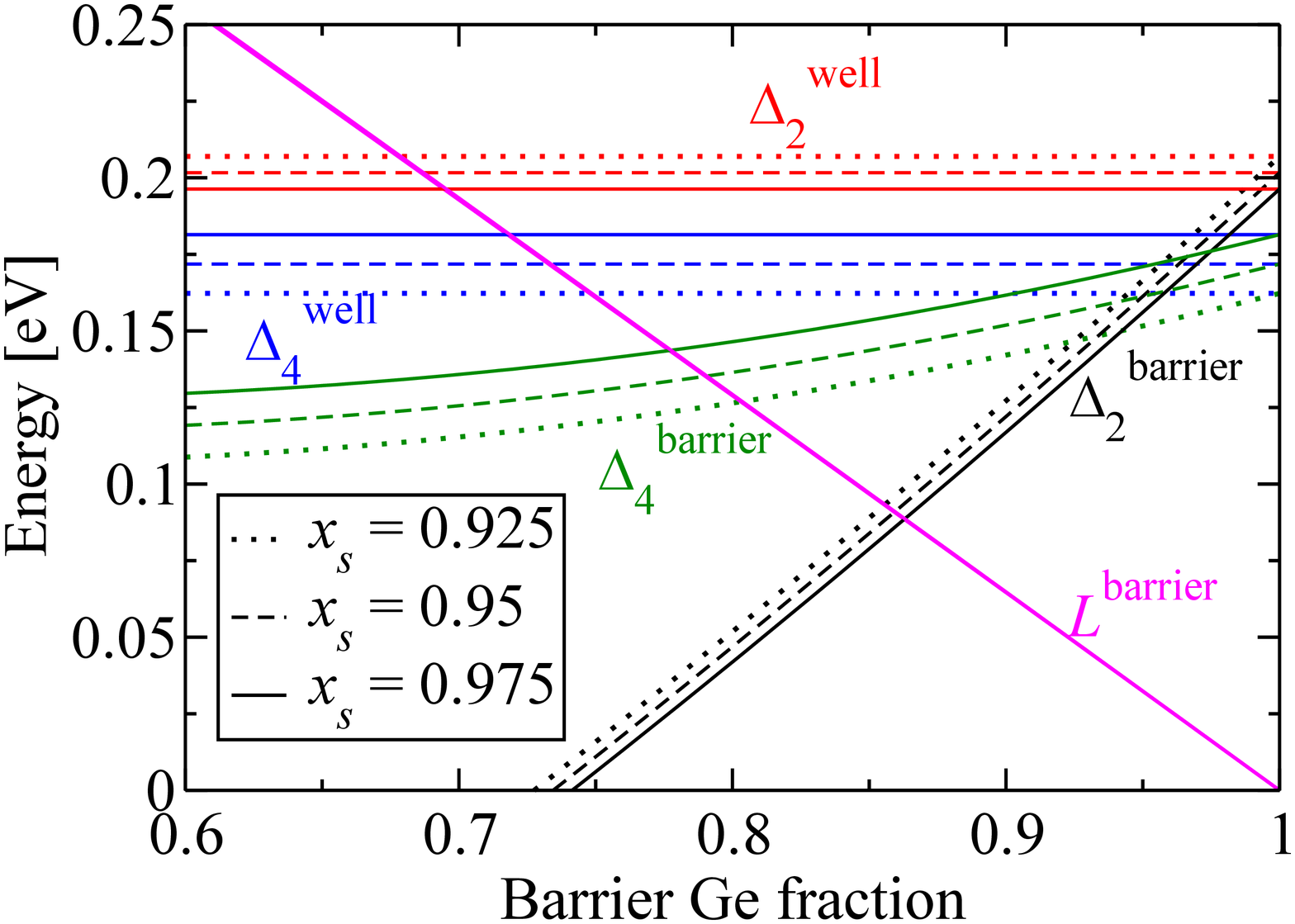}
		}
	\caption{\label{fig:usable_energy_ranges}(Color online) Conduction band 
	minima in the well and barrier regions of Si-based QWs relative to the
	bottom of the well.  Energies are plotted as a function of the barrier
	alloy composition, and results are shown for structures grown on various 
	Si$_{1-x_s}$Ge$_{x_s}$ virtual substrates.  The figures show band minima
	in QW heterostructures comprising \subref{subfig:usable-energy-Si001} 
	(001)-SiGe/Si/SiGe, \subref{subfig:usable-energy-Si111} 
	(111)-SiGe/Si/SiGe, and \subref{subfig:usable-energy-Ge001} 
	(001)-GeSi/Ge/GeSi.}
\end{figure*}

Fig.~\ref{fig:usable_energy_ranges} shows the conduction band minima in QWs that
consist of either a Si or Ge well surrounded by SiGe barriers.  The 
energies of the valley minima in the well and barriers are plotted as a
function of the barrier alloy composition.  In each case, the energies are
expressed relative to the bottom of the QW, and usable energy range is therefore 
given by the lowest line in the plot.

Fig.~\ref{fig:usable_energy_ranges}\subref{subfig:usable-energy-Si001} shows the
results for a (001)-oriented SiGe/Si/SiGe QW. Here, the $\Delta_2$ valleys form
the bottom of the well in the Si layer. The total depth of the QW is given by
the energy difference between the $\Delta_2$ minima in the barrier and the well.
As we shall see in section~\ref{scn:devices}, the barriers in (001)-Si/SiGe QCL
designs may need to be thinner than 1\,nm, owing to the large $\Delta_2$
quantization effective mass. It is, therefore, necessary to limit the barrier Ge
fraction to obtain a lower $\Delta_2$ confinement potential, and hence a
realistically wide barrier layer. We selected a Si$_{0.8}$Ge$_{0.2}$ alloy for
the barriers, which provides a $\Delta_2$ band offset of 95\,meV. A
$\sim$4.5\% Ge virtual substrate is required for mechanical stability in our
(001)-Si/SiGe QCL designs because the Si wells are considerably thicker than the
SiGe barrier layers. This induces only a relatively low uniaxial strain in the
QW layers and as a result the $\Delta_4$ minima in the barrier layers lie only
5\,meV above the bottom of the well. It is, therefore, impossible to avoid the
presence of $\Delta_4$ subbands within the energy range of interest for (001)
Si/SiGe QCL designs.

Fig.~\ref{fig:usable_energy_ranges}\subref{subfig:usable-energy-Si111} shows 
the calculated valley minima for (111)-oriented SiGe/Si/SiGe QWs. Here, the
$\Delta$ valleys are degenerate, and for most barrier compositions the usable
energy range is limited by the $\Delta$ valley offset at the Si/SiGe interface. 
The system is less sensitive to strain, and the maximum usable energy range of
185\,meV is obtained when the barriers have a Ge fraction of around 89\%.
However, our designs in section~\ref{scn:devices} use lower barriers
with a Si$_{0.4}$Ge$_{0.6}$ alloy composition, in order to obtain realistically
wide layer widths, as described above.  This composition provides a usable 
energy range of 90\,meV.

Fig.~\ref{fig:usable_energy_ranges}\subref{subfig:usable-energy-Ge001} shows 
the minima for (001)-oriented GeSi/Ge/GeSi QWs.  In structures with barrier
Ge fractions greater than around 75\%, the $L$ valleys form the bottom of the 
QW in the Ge layer.  However, for lower Ge alloys, the $\Delta_2$ valleys are 
lowest in energy. The maximum usable energy range of approximately
90\,meV is obtained when the barriers have a Ge fraction of around 0.85.  It is
worth noting that $\Delta_2$ states in the thin GeSi layers of QCLs will have
confinement energies well above the $\Delta_2$ band edge.  It may, therefore, be
possible to obtain a larger usable energy range by using a lower barrier Ge 
fraction.

\subsection{Self-consistent Poisson-Schr\"odinger solution}
\begin{figure}
	\subfloat[][]{
		\includegraphics*[width=8.6cm]{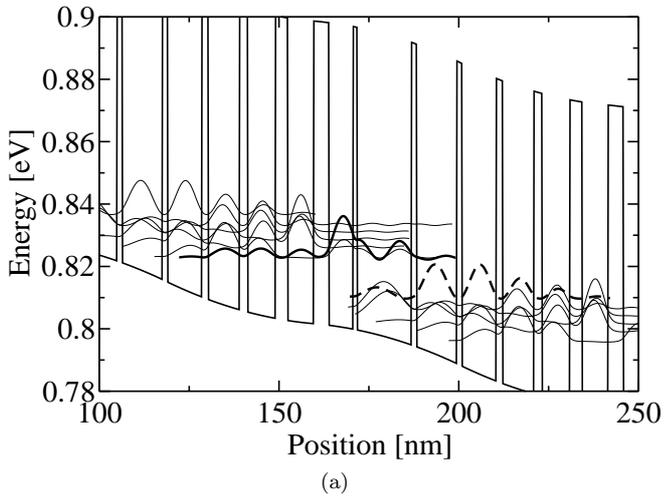}
		\label{subfig:wfplot_Ge_3THz}
	}\\
	\subfloat[][]{
		\includegraphics*[width=8.6cm]{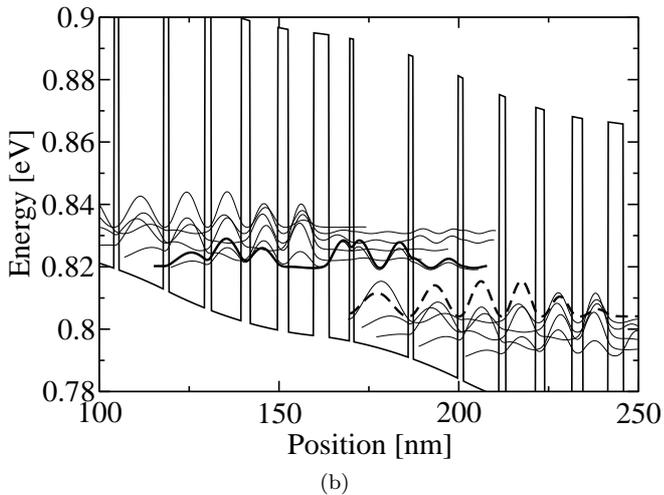}
		\label{subfig:wfplot_Ge_4THz}
	}

\caption{\label{fig:wfplot_Ge}Potential profile and electron probability
densities for optimized (001) Ge/GeSi QCL designs, with emission frequencies and
operating biases of \subref{subfig:wfplot_Ge_3THz} 3.0\,THz, 3.3\,kV/cm and
\subref{subfig:wfplot_Ge_4THz} 3.8\,THz, 3.5\,kV/cm.  The upper and lower 
subbands involved in the radiative transition are shown as solid-bold
and dashed-bold lines respectively.}
\end{figure}

\begin{table}[tb]
    \caption{\label{tbl:EffMassInterface}Quantization and two-dimensional
      density-of-states effective masses of conduction band valleys in (001) Si,
      (111) Si, and (001) Ge films, calculated as described in 
      Ref.~\onlinecite{JAPRahman2005}. Masses are expressed relative to the rest 
      mass of a free electron.}
    \begin{ruledtabular}
    \begin{tabular}{ccdd}
      Material system
          & Valley 
          & m_q 
          & m_d\\
      \hline
      \hline
      (001) Si
          & $\Delta_2$ 
          &0.916
          &0.19\\
      (111) Si 
          & $\Delta$
          & 0.26
          & 0.36\\
      (001) Ge 
          & $L$
          & 0.12
          & 0.30\\
    \end{tabular}
  \end{ruledtabular}
\end{table}

Self-consistent solutions of the one-dimensional time-independent 
Schr\"odinger equation and the Poisson equation were found for the 
structures considered in this work, using a similar approach to those in
Refs.~\onlinecite{NanotechCuratola2002} and \onlinecite{JAPRamMohan2004}.

The charge density over the length of the structure takes the form 
$\rho(z) = e[d(z) - \rho_e(z)]$, where $e$ is the electron charge, $d(z)$ is the
ionized donor profile, $\rho_e(z)$ is the electron density profile and $z$
is the spatial position.  It was assumed that all donors are ionized, and 
hence $d(z)$ is equal to the dopant distribution.  An initial estimate of the 
space-charge effect on the Hamiltonian was generated by solving the Poisson
equation for a
uniform electron distribution, in which $\rho_e(z) \approx N_{2D} / L_p$, where 
$N_{2D}$ is the sheet doping density across a structural period of length $L_p$.

A one-band parabolic effective mass approximation was used for the 
Schr\"odinger equation, which is 
justified by the small confinement energy, and the large energy separation from
other energy bands. The quantization and density-of-states effective masses
($m_q$ and $m_d$ respectively) were calculated for each material and crystal
orientation using the method in Ref.~\onlinecite{JAPRahman2005} and are shown in
table~\ref{tbl:EffMassInterface}.  

Intervalley mixing splits the $\Delta_2$ subbands in (001) Si/SiGe and the $L$ 
subbands in Ge/GeSi heterostructures.  It is, however, only possible to include 
this effect in complex heterostructures via computationally-expensive atomistic 
approaches such as tight-binding\cite{PRBVirgilio2009} or pseudopotential 
calculations.\cite{PRBValavanis2007} We have previously shown that the splitting
energies are small in structures wider than 2--3\,nm,\cite{PRBValavanis2007} and
have, therefore, omitted the effect in the present work.

Three periods of the conduction band potential profile were used in the 
Hamiltonian for the QCL simulations, with box boundary conditions.  This
introduced an unrealistic limit to the spatial extent of the wavefunctions which
were localized near to the boundaries.  To eliminate this effect, we 
replaced the Schr\"odinger equation solutions whose wavefunctions lay in the
left and right periods of the structure with translated copies of the central
period solutions.

Having solved the Schr\"odinger equation, the subband populations, $n_j$, were
calculated as described in the next section, and a new estimate of the charge
distribution was generated, using 
\begin{equation}
  \rho_{e}(z) = n_{\text{val}}\sum_jn_j |\psi_j(z)|^2,
\end{equation}
where $n_{\text{val}}$ is the valley degeneracy and $\psi_j$ is the 
wavefunction of state $j$.  The Poisson and Schr\"odinger equations were then 
solved iteratively to find the self-consistent solutions.

\section{Carrier transport model}
\label{scn:transport}
\begin{figure}
	\subfloat[][]{
		\includegraphics*[width=8.6cm]{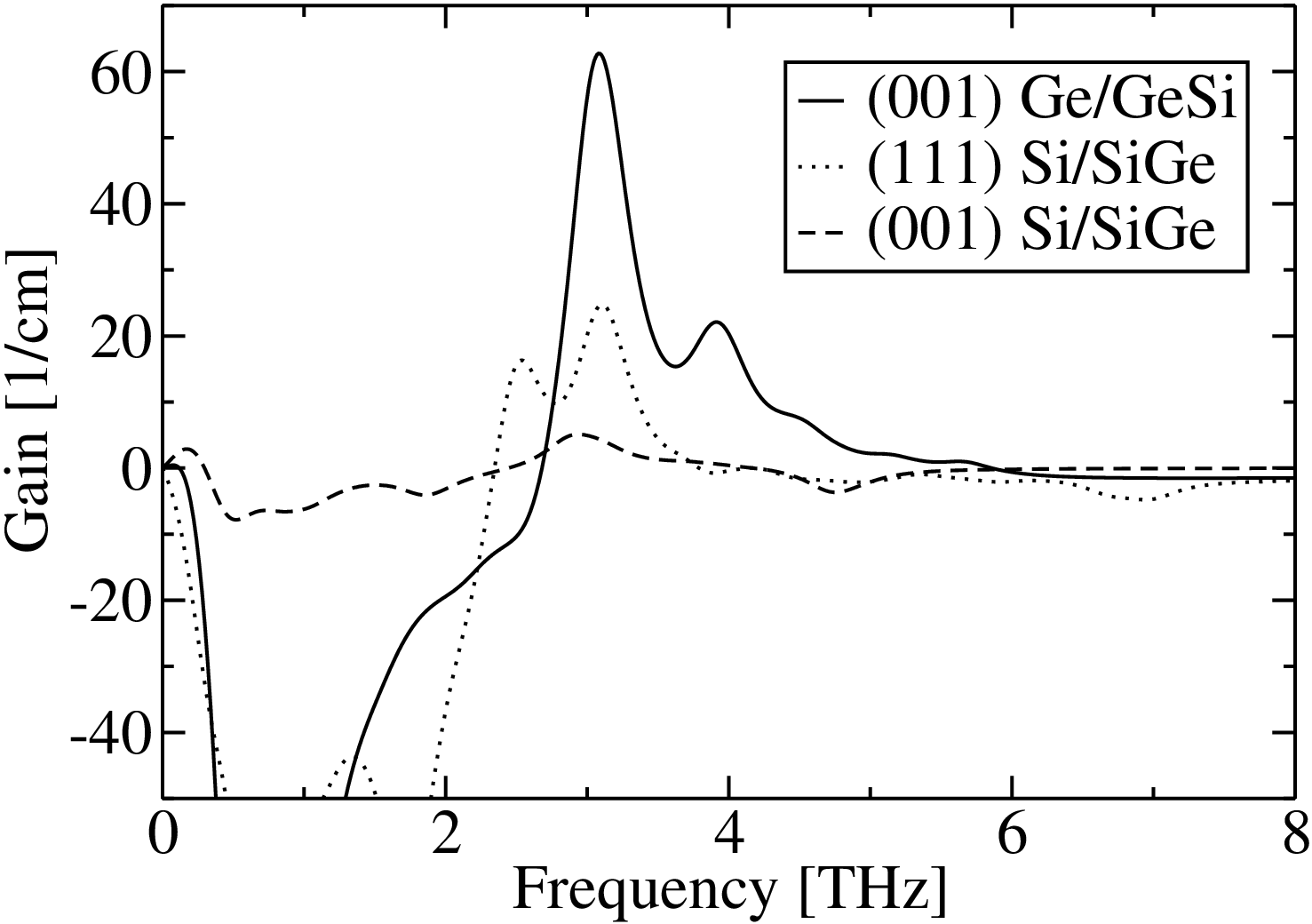}
		\label{subfig:gain-spectra-3THz}
	}\\
	\subfloat[][]{
		\includegraphics*[width=8.6cm]{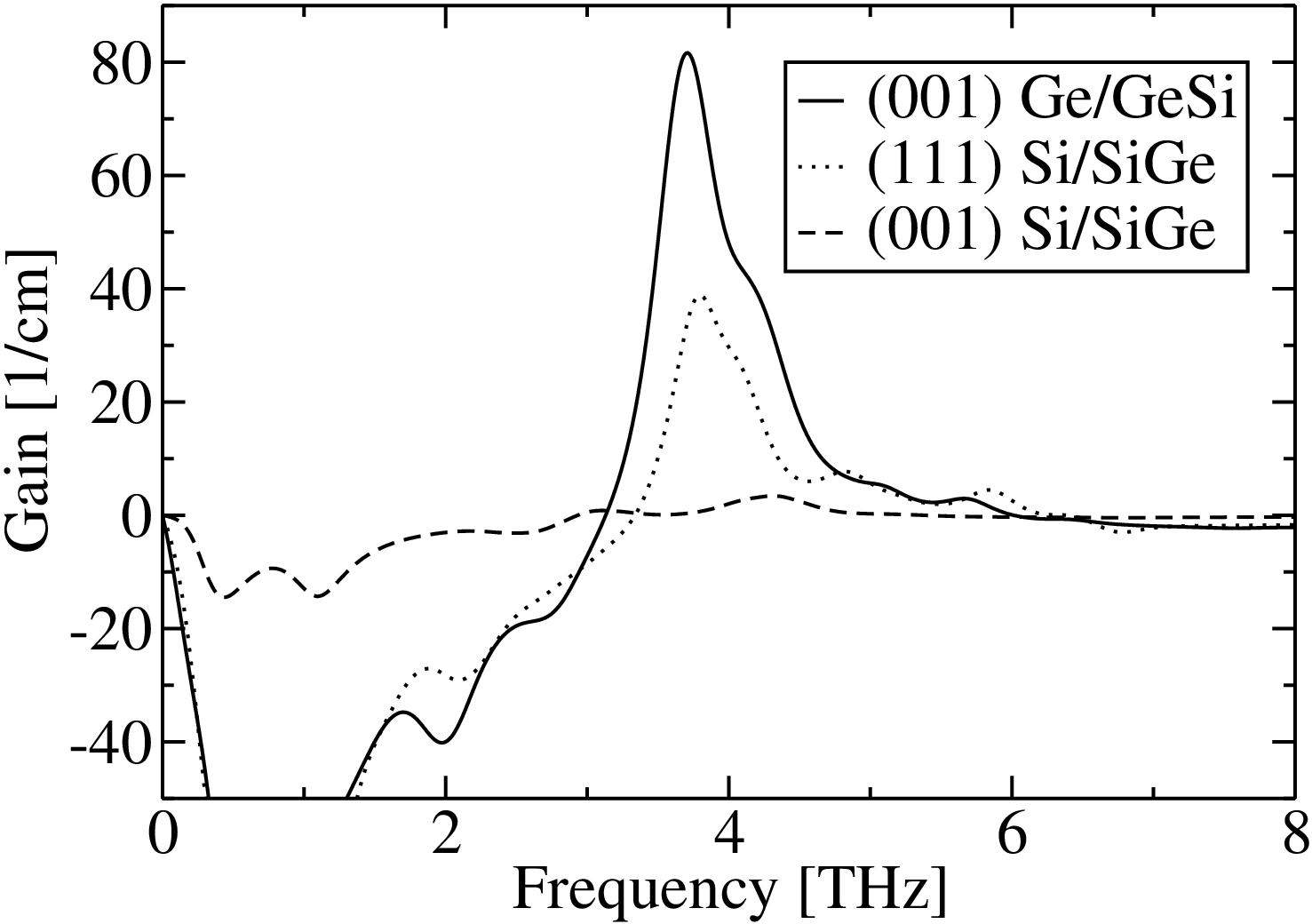}
		\label{subfig:gain-spectra-4THz}
	}
	\caption{\label{fig:gain-spectra}Gain spectra for devices emitting near 
	\subref{subfig:gain-spectra-3THz} 3\,THz and 
	\subref{subfig:gain-spectra-4THz} 4\,THz.}
\end{figure}
We used a computationally-efficient scattering calculation 
and a semi-classical approach to electron transport in QCLs as described in
our previous work.\cite{PRBValavanis2008_2, PRBValavanis2008}
Similar Boltzmann or rate equation based models have previously yielded good
agreement with experimental data for III--V mid-infrared\cite{JAPJovanovic2006} 
and THz QCLs.\cite{NatureKohler2002}  Indeed, our carrier-transport model has
been shown to calculate the current density and maximum operating temperature 
quite accurately for a 4.4\,THz, 7-well chirped-superlattice GaAs/AlGaAs QCL
that is similar in structure to the devices considered in
section~\ref{scn:devices}.\cite{SSTIndjin2005} Although recent work on III--V
QCLs has focused on coherent transport effects, using nonequilibrium Green's
functions\cite{PRBLee2002} or density matrix\cite{LPRevScalari2009,
JAPCallebaut2005,PRBDupont2010,PRBKumar2009,PRBGordon2009,PRBWang2010}
approaches, the semi-classical approach combines the computational speed and
flexibility required for extensive exploration of the parameter space of
possible device specifications. We have assumed that coherent transport effects
are less significant in the bound--to--continuum devices considered in this work
than in resonant-phonon QCLs, owing to the reduced thickness of the injection
barriers through which electrons tunnel into the active region. Furthermore, the
absence of resonant LO-phonon scattering may lead to longer dephasing times for
coherent transport in group-IV materials than in III--V materials.

As in our previous
work,\cite{PRBValavanis2008_2, PRBValavanis2008} our model includes elastic
intravalley scattering due to interface roughness (allowing arbitrary interface
geometries),\cite{PRBValavanis2008} alloy disorder,\cite{PRBQuang2007,
PRLMurphy-Armando2006} ionized impurities,\cite{JAPUnuma2003} electron--electron
interactions\cite{JAPSmet1996} and deformation potential scattering for
electron--acoustic phonon interactions.\cite{PRBFischetti1993} Intravalley
optical phonon interactions are forbidden in $\Delta$ valleys due to the
symmetry of the system,\cite{Ridley1997} but were included for $L$ valleys via a
zero-order deformation potential model.\cite{PRBFischetti1993,
IEEEFischetti1991}

Intervalley phonon scattering was also described using the zero-order
deformation potential model, with the rates multiplied by the number of equivalent 
destination valleys.  In $L\to{}L$ scattering, all three destination valleys
are degenerate, and separated by a wavevector of the same magnitude.  A
phenomenological approach, described in Ref.~\onlinecite{PRBJacoboni1981}, was
used to describe the $L\to{}L$ interactions by treating the combined scattering 
from all phonon branches as a single interaction. 
$\Delta\to\Delta$ scattering interactions are categorized as either $g$ type, in
which the destination valley lies on the same crystallographic axes as the
source, or $f$ type in which the destination valleys lie on a different axis.
Phonon energies and deformation potentials for $g$ and $f$ interactions with
longitudinal/transverse optical (LO/TO) and acoustic (LA/TA) phonon branches
were taken from Ref.~\onlinecite{JAPDollfus1997}. The high-energy $g$-LO, $f$-LA
and $f$-TO phonon interactions were determined using the zero-order deformation
potential model. The lower-energy $g$-TA, $g$-LA, and $f$-TA
interactions have no zero-order component in their deformation potential, owing
to symmetry selection rules, and were instead determined using a first-order
model.\cite{PRBFerry1976, PRBMonsef2002, *PRBMonsef2002Errata} The bandstructure
calculations in section~\ref{scn:bandstructure} show that $\Delta_4$ quantum
wells correspond to $\Delta_2$ barriers in (001) Si/SiGe heterostructures. The
small spatial overlap of wavefunctions leads to very small $\Delta_2\to\Delta_4$
scattering matrix elements, and $f$ transitions were therefore omitted in our
model of (001) devices as a first-approximation.

The steady-state populations $n_i$ for each subband were found using a 
rate-equation approach.\cite{JAPJovanovic2006}  Intrasubband scattering rates
were typically calculated to be an order of magnitude faster than intersubband 
scattering.  It was therefore assumed that electrons settle between intersubband
scattering events to a quasi-thermal Fermi-Dirac distribution.
The distribution for each subband was described by a quasi-Fermi energy
$E_{F,i}$ and a global electron temperature $T_e$ and the total subband 
population was found using
\begin{equation}
    n_i = \rho^{2D} k_B T_e \ln\left[1+\me^{\frac{E_{F,i}(T_e)}{k_BT_e}}\right],
\end{equation}
where $\rho^{2D} = m_d/(\pi\hbar^2)$ is the two-dimensional density-of-states.

A root-finding approach was used to determine the steady-state electron 
temperature at which no net gain or loss of kinetic energy occurred within the
QCL, using the expression\cite{JAPJovanovic2006}
\begin{equation}
	\frac{\dif{E_k}}{\dif{t}} = \sum_f\sum_i n_i(T_e) E_{if} \overline{W}_{if}(T_e) = 0.
\end{equation}
Here, $\overline{W}_{if}$ is the average intersubband scattering rate between
a pair of subbands $i$ and $j$, summed over all scattering processes, and 
$E_{if}$ is the energy difference between the subband minima.  In the case of 
inelastic processes, the transition energy was modified as $E_{if}\to
E_{if}\pm\hbar\omega_q$ to account for the absorption or emission of a phonon
with energy $\hbar\omega_q$.

The current density was estimated by considering the average scattering rates
and the change in electron position for all intersubband 
transitions,\cite{PRBValavanis2008_2}
\begin{equation}
	J = \frac{e}{L_p}\sum_i n_in_{\text{val}}^i\sum_f \left(\langle z \rangle_f - \langle z \rangle_i\right)\overline{W}_{if},
\end{equation}
where $L_p$ is the length of a structural period of the QCL, $n_{\text{val}}$ is
the number of equivalent initial valleys and $\langle z \rangle$ is the
expectation position for an electron in a given subband.

The optical gain per unit length was calculated using $G(\omega) =
\sigma(\omega)/(\varepsilon_0 cn_r)$,\cite{Davies1998, JAPJovanovic2006} where
$n_r$ is the real part of the refractive index of the active region stack and
$\sigma$ is the real part of the optical conductivity. This is given by
\begin{equation}
\sigma(\omega) 
= \frac{\pi e^2}{2(m_qm_d^2)^{\frac{1}{3}}L_p}
\sum_{i,j}f_{ji}n_in_{\text{val}}^i\operatorname{sgn}(E_{ij})L_{ij}(\omega),
\end{equation}
where $L_{ij}(\cdot)$ is a lineshape function and $\operatorname{sgn}(\cdot)$
represents the sign-function. The oscillator strength is given by
\begin{equation}
	f_{ji} = \frac{2(m_qm_d^2)^{\frac{1}{3}}}{\hbar}\omega_{ij}|z_{ij}|^2,
\end{equation}
where, $z_{ij}=\langle j|z|i\rangle$ is the dipole matrix element.  A Lorentzian
lineshape was assumed, with a linewidth of 
2\,meV, as is typical for the lasing transition in GaAs based
THz QCLs.\cite{APLWalther2006, APLScalari2005}

\section{Device designs}
\label{scn:devices}
\begin{figure}
\includegraphics*[width=8.6cm]{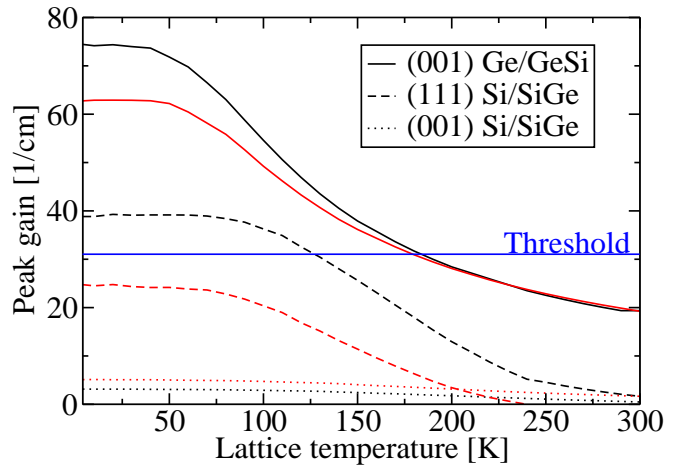}
\caption{(Color online) Peak gain as a function of lattice temperature for 
devices emitting near 3\,THz (red lines) and 4\,THz (black lines).  The blue
line represents an indicative figure of 31\,cm$^{-1}$ for the threshold gain.}
\label{fig:Gpk-T-combined}
\end{figure}
Equivalent 3 and 4\,THz QCLs were designed for each of the three material 
configurations, in order to simulate the relative performance of each system.
To obtain a fair comparison between the materials, all designs were generated 
using an automated process. In principle, it would have been possible to
generate slightly better designs than those presented in this section by
manually adjusting the device structures at the start and end of the automated
design process. However, we chose not to apply any manual design optimization
in this work, in order to ensure that a consistent, reproducible and systematic 
exploration of the design parameter space was used for each material 
configuration.

Pure Si or Ge was
used as the well material in order to prevent depopulation of the upper laser
level via alloy disorder scattering. For each material configuration, a SiGe
alloy was chosen for the barriers to maximize the usable energy range
(as described in section~\ref{scn:bandstructure}), without introducing 
mechanical instability.

A recent seven-well bound--to--continuum (BTC) (001) Ge/GeSi design was selected
as a template for all the designs considered.  This device gives a large 
simulated gain ($\sim{50}$\,cm$^{-1}$) at 3.5\,THz, with a threshold current 
density around 300\,A/cm$^2$, an operating bias of 
3.6\,kV/cm,\cite{APLLever2009} and a maximum operating temperature of 
136\,K.\cite{Thesis_Valavanis2009}
In this structure, doping was spread evenly over four wells and three barriers 
in the injector region of each QCL period, with a total sheet
density of $8 \times 10^{10}\,\mathrm{cm^{-2}}$.  Complete donor ionization was
assumed.  A similar seven-well BTC
device has been demonstrated in the GaAs/AlGaAs material system, with an
emission frequency of 3.66\,THz at an operating bias of 4.15\,kV/cm and with a
threshold current density of $\sim$200\,A/cm at low operating temperatures. This
GaAs/AlGaAs device was shown to have a maximum operating temperature of
116\,K---the highest reported for a BTC THz QCL.\cite{APLScalari2005}

An automated design algorithm\cite{PhysStatSol_Ko_2010} was used to generate
a pair of new QCL designs from the Ge/GeSi QCL template, with emission 
frequencies of
3 and 4\,THz.  In this method, the gain spectrum was calculated using a range of
external electric fields from 3 to 10\,kV/cm. The thickness of each barrier and
well was adjusted sequentially to maximize the gain in a frequency window of
$\pm$200\,GHz around the desired emission frequency. The entire process was
iterated until the algorithm converged on an optimal layer structure and field.
The optimal virtual substrate composition was calculated for each
device,\cite{Harrison2005} to ensure zero net stress across each period of the
QCL.

Equivalent 3 and 4\,THz QCL designs were generated using (001) and 
(111)-oriented Si/SiGe configurations. In each case, the thickness $l$ of each
layer in the template QCL structure was transformed according to $l' =
l\sqrt{m_q/m_q'}$,\cite{JOAValavanis2009} where the prime notation denotes
parameters of the Si/SiGe system in the appropriate orientation. This
transformation yields subband spacings approximately equal to those in the
original Ge/GeSi design template. The automated design algorithm was then
applied as described above.

\begin{table*}[tb]
    \caption{\label{tbl:Designs}QCL design parameters for each of the devices
             designed in this work, where $f_0$ is the emission frequency in
             THz, $x_s$ denotes the virtual substrate Ge fraction and $F$ is the
             operating bias in kV/cm. Bold text in the layer structure
             represents barriers, while normal weighted text represents wells.
             Doped layers are underlined.} 
  \begin{ruledtabular}
    \begin{tabular}{ldldd}
      Material & f_0 & Layers [nm] & x_s & F\\
      \hline
      \hline
      (001) Si/Si$_{0.8}$Ge$_{0.2}$ & 2.9 & 
      		2.1/\textbf{0.8}/6.4/\textbf{0.7}/3.5/\textbf{1.0}/\underline{3.3}/\textbf{\underline{1.2}}/\underline{3.1}/\textbf{\underline{1.4}}/\underline{3.0}/\textbf{\underline{1.4}}/\underline{2.9}/\textbf{1.6} & 0.048 & 7.1 \\
      (111) Si/Si$_{0.4}$Ge$_{0.6}$ & 3.1 & 
      		3.2/\textbf{1.0}/9.2/\textbf{0.8}/5.0/\textbf{1.5}/\underline{4.5}/\textbf{\underline{1.9}}/\underline{4.4}/\textbf{\underline{2.3}}/\underline{4.4}/\textbf{\underline{2.4}}/\underline{4.2}/\textbf{3.3} & 0.146 & 6.9 \\
      (001) Ge/Ge$_{0.85}$Si$_{0.15}$ & 3.0 & 
      		6.7/\textbf{1.2}/15.1/\textbf{1.4}/11.1/\textbf{1.5}/\underline{9.5}/\textbf{\underline{1.8}}/\underline{8.7}/\textbf{\underline{2.3}}/\underline{7.7}/\textbf{\underline{3.5}}/\underline{7.1}/\textbf{4.3} & 0.969 & 3.3 \\
      	\hline
      	\hline
      (001) Si/Si$_{0.8}$Ge$_{0.2}$ & 4.1 & 
      		2.2/\textbf{0.7}/6.1/\textbf{0.8}/4.3/\textbf{1.0}/\underline{3.1}/\textbf{\underline{1.0}}/\underline{3.1}/\textbf{\underline{1.2}}/\underline{3.0}/\textbf{\underline{1.4}}/\underline{2.7}/\textbf{1.5} & 0.045 & 7.3 \\
      (111) Si/Si$_{0.4}$Ge$_{0.6}$ & 4.0 & 
      		3.1/\textbf{1.0}/8.7/\textbf{1.3}/5.3/\textbf{1.6}/\underline{5.0}/\textbf{\underline{2.0}}/\underline{4.6}/\textbf{\underline{2.1}}/\underline{4.0}/\textbf{\underline{2.4}}/\underline{4.0}/\textbf{3.7} & 0.154 & 6.9 \\
      (001) Ge/Ge$_{0.85}$Si$_{0.15}$ & 3.8 & 
      		5.8/\textbf{1.0}/15.3/\textbf{1.4}/12.3/\textbf{1.6}/\underline{9.9}/\textbf{\underline{1.9}}/\underline{8.3}/\textbf{\underline{2.4}}/\underline{7.8}/\textbf{\underline{2.9}}/\underline{7.0}/\textbf{4.3} & 0.970 & 3.5 \\
    \end{tabular}
    \end{ruledtabular}
\end{table*}

Parameters for each of the final QCL designs are summarized in
table~\ref{tbl:Designs}.  It can be seen that the Si/SiGe device designs 
generally require thinner layers than the Ge/GeSi designs, owing to the 
difference in effective mass.  Epitaxial growth of QCLs in Si/SiGe may, 
therefore, be more challenging.  The total length of an active region period
is also lower in Ge/GeSi designs than in Si/SiGe, which leads to a lower
operating bias.
The bandstructure and electron probability densities for the Ge/GeSi designs
are plotted in Fig.~\ref{fig:wfplot_Ge}.

\section{Simulated device performance}
\label{scn:results}
\begin{figure}
\includegraphics*[width=8.6cm]{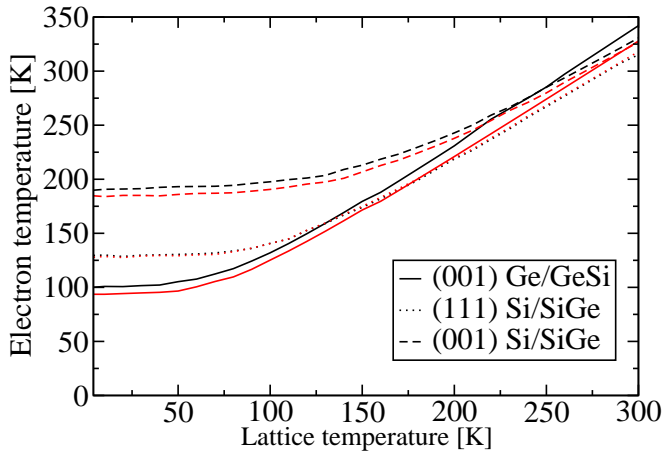}
\caption{(Color online) Relationship between electron temperature and lattice temperature for
devices emitting near 3\,THz (red lines) and 4\,THz (black lines).}
\label{fig:Te_vs_T-combined}
\end{figure}

The simulated gain spectra for all devices, operating at their respective
design biases are shown in Fig.~\ref{fig:gain-spectra}.  It can be seen that
the Ge/GeSi designs yield the highest gain at both frequencies, with peaks of
63 and 82\,cm$^{-1}$ for the 3 and 4\,THz designs respectively.  The
peak gains for the Si/SiGe designs are significantly lower: 25 and
38\,cm$^{-1}$ for the (111)-oriented devices, and 5 and 3\,cm$^{-1}$ for
the (001)-oriented devices at 3 and 4\,THz respectively.  We have previously
calculated a threshold gain of 31\,cm$^{-1}$ for a 15-$\mu$m-thick (001) Si/SiGe
QCL active region in a copper double-metal waveguide 
structure.\cite{Thesis_Valavanis2009}  By taking this threshold as an 
indicative figure, we predict that net gain is achievable for both Ge/GeSi 
devices, and for the 4\,THz (111) Si/SiGe device.

The peak gains in the spectra decrease as the lattice temperature increases, 
as shown in Fig.~\ref{fig:Gpk-T-combined}, owing to the reduction in population
inversion.  This is caused by a number of thermal effects, including electron
leakage from the upper laser level via phonon emission, and by thermal 
backfilling of the lower laser level.  Net gain is 
predicted for the Ge/GeSi devices up to lattice temperatures of 179 and 184\,K 
for 3 and 4\,THz emission respectively. The 4\,THz (111) Si/SiGe device is
predicted to yield net gain up to a lattice temperature of 127\,K.

\begin{figure}
\includegraphics*[width=8.6cm]{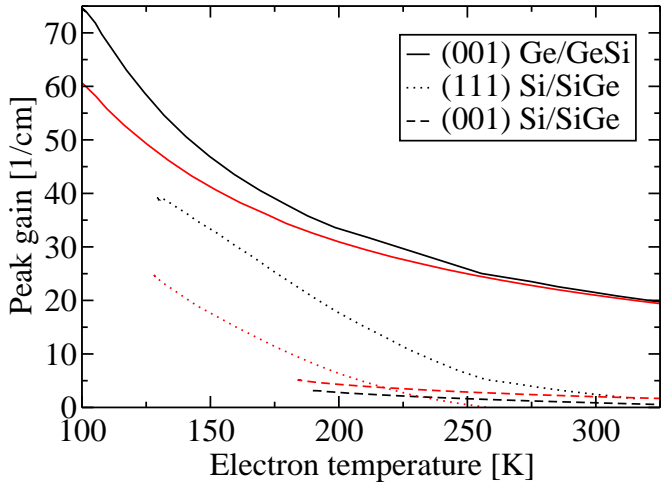}
\caption{(Color online) Relationship between peak gain and electron temperature for
devices emitting near 3\,THz (red lines) and 4\,THz (black lines).}
\label{fig:Gpk_vs_Te-combined}
\end{figure}

The simulated temperature of the electron distribution $T_e$ is plotted as a
function of lattice temperature $T$ in Fig.~\ref{fig:Te_vs_T-combined}.  
At high lattice temperatures, $T_e$ is a linear function of $T$ and is 
approximately independent of bias.  At low lattice temperatures, however,  
$T_e$ is determined principally by the applied electric field.  In the case of 
(001) Si/SiGe devices, the bias is relatively large ($>7$\,kV/cm), and electrons
therefore scatter preferentially into high-energy states. This yields high 
steady-state
electron temperatures of 184 and 189\,K for emission at 3 and 4\,THz
respectively at a lattice temperature of 4\,K. The electric fields are lower in
(111) Si/SiGe and (001) Ge/GeSi devices, owing to the greater lengths of the
active regions. This leads to correspondingly lower electron temperatures of 127
and 129\,K for (111) Si/SiGe devices, and 93 and 100\,K for Ge/GeSi
devices emitting at 3 and 4\,THz respectively.
The effect of thermal excitation upon device
performance is illustrated in Fig.~\ref{fig:Gpk_vs_Te-combined}. It can be seen
that the gain decreases monotonically as electron temperature increases, owing
to the thermal backfilling of the lower laser level. Ge/GeSi devices are able to
operate with the lowest electron temperatures, and hence achieve the
highest peak gains.

The current density was calculated at the design bias for each of the devices.
Current densities of 270 and 380\,A\,cm$^{-2}$ were predicted at the design bias 
for Ge/GeSi devices operating at 3 and 4\,THz respectively.  In Si/SiGe, current densities were 
calculated as 430 and 460\,A\,cm$^{-2}$ for the (111)-oriented devices and
210 and 240\,A\,cm$^{-2}$ for the (001)-oriented devices at 3 and 4\,THz respectively.
The low operating currents in (001) Si/SiGe devices were due to the very low 
scattering rates, which result from the high $\Delta_2$ valley effective mass.
The ratio of peak gain to current density was calculated as a figure of merit
for each device at its design bias. Ge/GeSi devices were found to have the
highest values (240 and 210\,cm/kA) followed by (111) Si/SiGe (57 and
84\,cm/kA), and (001) Si/SiGe (25 and 14\,cm/kA) at 3 and 4\,THz
respectively.  We should note that our simulations of
(001) Si/SiGe QCLs do not include $\Delta_2\to\Delta_4$ intervalley scattering
events, which would further degrade the predicted performance. However, as these
structures already appear to be poor candidates for laser design, a more
comprehensive transport model was considered unnecessary.
Threshold current densities were calculated at $T=4\,K$
as 440, 210, and 330\,A\,cm$^{-2}$ for the 4\,THz (111)-Si/SiGe, 3\,THz Ge/GeSi and
4\,THz Ge/GeSi devices respectively. 

\section{Conclusion}
We have presented a comparison between the simulated performance of Si-based 
QCLs using the (001) Ge/GeSi, (111) Si/SiGe, and (001) Si/SiGe material
configurations.  A semi-automated design optimization algorithm was used, in
order to provide a fair comparison between equivalent designs.  Our results
show that (001) Ge/GeSi is the most promising system for development of a
Si-based QCL.  Firstly, the 
bandstructure calculations in section~\ref{scn:bandstructure} show that the
(001) Ge/GeSi and (111) Si/SiGe systems offer a $\gtrsim$90\,meV energy range
for QCL design, compared with only $\sim$5\,meV in (001) Si/SiGe systems, owing 
to the large energy separation between conduction band minima. This reduces the
probability of current-leakage via intervalley scattering, and allows a wider
range of emission frequencies to be targeted.
Secondly, the low $L$ valley effective mass was found to yield a relatively long
period length for the QCL active region. This reduces the operating electric
field, and hence the current density and the temperature of the electron
distribution.  Net gain was predicted for both of the Ge/GeSi devices, but only 
one of the four optimized Si/SiGe devices. Ge/GeSi bound--to--continuum QCLs
were predicted to operate up to temperatures of 179 and 184\,K at 3 and 4\,THz
respectively, while the 4\,THz (111) Si/SiGe device was predicted to operate
up to 127\,K.  These figures may potentially be improved via waveguide design
optimization to minimize losses, or through the use of a resonant-phonon active 
region design.
Nevertheless, the predicted values exceed the highest-recorded operating
temperature of 116\,K for a 3.66\,THz seven-well III--V BTC
device.\cite{APLScalari2005}

\begin{acknowledgments}
This work was supported by EPSRC Doctoral Training Allowance funding. The
authors are grateful to Jonathan Cooper, University of Leeds and Douglas Paul, 
University of Glasgow for useful discussions.
\end{acknowledgments}


\begin{thebibliography}{10}%
\makeatletter
\providecommand \@ifxundefined [1]{%
 \ifx #1\undefined \expandafter \@firstoftwo
 \else \expandafter \@secondoftwo
\fi
}%
\providecommand \@ifnum [1]{%
 \ifnum #1\expandafter \@firstoftwo
 \else \expandafter \@secondoftwo
\fi
}%
\providecommand \enquote [1]{``#1''}%
\providecommand \bibnamefont  [1]{#1}%
\providecommand \bibfnamefont [1]{#1}%
\providecommand \citenamefont [1]{#1}%
\providecommand\href[0]{\@sanitize\@href}%
\providecommand\@href[1]{\endgroup\@@startlink{#1}\endgroup\@@href}%
\providecommand\@@href[1]{#1\@@endlink}%
\providecommand \@sanitize [0]{\begingroup\catcode`\&12\catcode`\#12\relax}%
\@ifxundefined \pdfoutput {\@firstoftwo}{%
 \@ifnum{\z@=\pdfoutput}{\@firstoftwo}{\@secondoftwo}%
}{%
 \providecommand\@@startlink[1]{\leavevmode}%
 \providecommand\@@endlink[0]{}%
}{%
 \providecommand\@@startlink[1]{%
  \leavevmode
  \pdfstartlink
   attr{/Border[0 0 1 ]/H/I/C[0 1 1]}%
   user{/Subtype/Link/A<</Type/Action/S/URI/URI(#1)>>}%
  \relax
 }%
 \providecommand\@@endlink[0]{\pdfendlink}%
}%
\providecommand \url  [0]{\begingroup\@sanitize \@url }%
\providecommand \@url [1]{\endgroup\@href {#1}{\urlprefix}}%
\providecommand \urlprefix [0]{URL }%
\providecommand \Eprint[0]{\href }%
\@ifxundefined \urlstyle {%
  \providecommand \doi [1]{doi:\discretionary{}{}{}#1}%
}{%
  \providecommand \doi [0]{doi:\discretionary{}{}{}\begingroup
  \urlstyle{rm}\Url }%
}%
\providecommand \doibase [0]{http://dx.doi.org/}%
\providecommand \Doi[1]{\href{\doibase#1}}%
\providecommand \bibAnnote [3]{%
  \BibitemShut{#1}%
  \begin{quotation}\noindent
    \textsc{Key:}\ #2\\\textsc{Annotation:}\ #3%
  \end{quotation}%
}%
\providecommand \bibAnnoteFile [2]{%
  \IfFileExists{#2}{\bibAnnote {#1} {#2} {\input{#2}}}{}%
}%
\providecommand \typeout [0]{\immediate \write \m@ne }%
\providecommand \selectlanguage [0]{\@gobble}%
\providecommand \bibinfo [0]{\@secondoftwo}%
\providecommand \bibfield [0]{\@secondoftwo}%
\providecommand \translation [1]{[#1]}%
\providecommand \BibitemOpen[0]{}%
\providecommand \bibitemStop [0]{}%
\providecommand \bibitemNoStop [0]{.\EOS\space}%
\providecommand \EOS [0]{\spacefactor3000\relax}%
\providecommand \BibitemShut [1]{\csname bibitem#1\endcsname}%
\bibitem{NatureKohler2002}%
  \BibitemOpen
  \bibfield{author}{%
  \bibinfo {author} {\bibfnamefont{R.}~\bibnamefont{K\"ohler}}, \bibinfo
  {author} {\bibfnamefont{A.}~\bibnamefont{Tredicucci}}, \bibinfo {author}
  {\bibfnamefont{F.}~\bibnamefont{Beltram}}, \bibinfo {author}
  {\bibfnamefont{H.~E.}\ \bibnamefont{Beere}}, \bibinfo {author}
  {\bibfnamefont{E.~H.}\ \bibnamefont{Linfield}}, \bibinfo {author}
  {\bibfnamefont{A.~G.}\ \bibnamefont{Davies}}, \bibinfo {author}
  {\bibfnamefont{D.~A.}\ \bibnamefont{Ritchie}}, \bibinfo {author}
  {\bibfnamefont{R.~C.}\ \bibnamefont{Iotti}},\ and\ \bibinfo {author}
  {\bibfnamefont{F.}~\bibnamefont{Rossi}},\ }%
  \bibfield{journal}{%
  \Doi{10.1038/417156a}{\bibinfo {journal} {Nature}}\ }%
  \textbf{\bibinfo {volume} {417}},\ \bibinfo {pages} {156} (\bibinfo {year}
  {2002})%
  \bibAnnoteFile{NoStop}{NatureKohler2002}%
\bibitem{NatPhotonicsWilliams2007}%
  \BibitemOpen
  \bibfield{author}{%
  \bibinfo {author} {\bibfnamefont{B.~S.}\ \bibnamefont{Williams}},\ }%
  \bibfield{journal}{%
  \Doi{10.1038/nphoton.2007.166}{\bibinfo {journal} {Nat. Photonics}}\ }%
  \textbf{\bibinfo {volume} {1}},\ \bibinfo {pages} {517} (\bibinfo {year}
  {2007})%
  \bibAnnoteFile{NoStop}{NatPhotonicsWilliams2007}%
\bibitem{NatureHolland1998}%
  \BibitemOpen
  \bibfield{author}{%
  \bibinfo {author} {\bibfnamefont{W.~S.}\ \bibnamefont{Holland}}, \bibinfo
  {author} {\bibfnamefont{J.~S.}\ \bibnamefont{Greaves}}, \bibinfo {author}
  {\bibfnamefont{B.}~\bibnamefont{Zuckerman}}, \bibinfo {author}
  {\bibfnamefont{R.~A.}\ \bibnamefont{Webb}}, \bibinfo {author}
  {\bibfnamefont{C.}~\bibnamefont{McCarthy}}, \bibinfo {author}
  {\bibfnamefont{I.~M.}\ \bibnamefont{Coulson}}, \bibinfo {author}
  {\bibfnamefont{D.~M.}\ \bibnamefont{Walther}}, \bibinfo {author}
  {\bibfnamefont{W.~R.~F.}\ \bibnamefont{Dent}}, \bibinfo {author}
  {\bibfnamefont{W.~K.}\ \bibnamefont{Gear}},\ and\ \bibinfo {author}
  {\bibfnamefont{I.}~\bibnamefont{Robson}},\ }%
  \bibfield{journal}{%
  \Doi{10.1038/33874}{\bibinfo {journal} {Nature}}\ }%
  \textbf{\bibinfo {volume} {392}},\ \bibinfo {pages} {788} (\bibinfo {year}
  {1998})%
  \bibAnnoteFile{NoStop}{NatureHolland1998}%
\bibitem{OptExpLoffler2001}%
  \BibitemOpen
  \bibfield{author}{%
  \bibinfo {author} {\bibfnamefont{T.}~\bibnamefont{L{\"o}ffler}}, \bibinfo
  {author} {\bibfnamefont{T.}~\bibnamefont{Bauer}}, \bibinfo {author}
  {\bibfnamefont{K.}~\bibnamefont{Siebert}}, \bibinfo {author}
  {\bibfnamefont{H.}~\bibnamefont{Roskos}}, \bibinfo {author}
  {\bibfnamefont{A.}~\bibnamefont{Fitzgerald}},\ and\ \bibinfo {author}
  {\bibfnamefont{S.}~\bibnamefont{Czasch}},\ }%
  \bibfield{journal}{%
  \Doi{10.1364/OE.9.000616}{\bibinfo {journal} {Opt. Express}}\ }%
  \textbf{\bibinfo {volume} {9}},\ \bibinfo {pages} {616} (\bibinfo {year}
  {2001})%
  \bibAnnoteFile{NoStop}{OptExpLoffler2001}%
\bibitem{PRBCalifano2007}%
  \BibitemOpen
  \bibfield{author}{%
  \bibinfo {author} {\bibfnamefont{M.}~\bibnamefont{Califano}}, \bibinfo
  {author} {\bibfnamefont{N.~Q.}\ \bibnamefont{Vinh}}, \bibinfo {author}
  {\bibfnamefont{P.~J.}\ \bibnamefont{Phillips}}, \bibinfo {author}
  {\bibfnamefont{Z.}~\bibnamefont{Ikonic}}, \bibinfo {author}
  {\bibfnamefont{R.~W.}\ \bibnamefont{Kelsall}}, \bibinfo {author}
  {\bibfnamefont{P.}~\bibnamefont{Harrison}}, \bibinfo {author}
  {\bibfnamefont{C.~R.}\ \bibnamefont{Pidgeon}}, \bibinfo {author}
  {\bibfnamefont{B.~N.}\ \bibnamefont{Murdin}}, \bibinfo {author}
  {\bibfnamefont{D.~J.}\ \bibnamefont{Paul}}, \bibinfo {author}
  {\bibfnamefont{P.}~\bibnamefont{Townsend}}, \bibinfo {author}
  {\bibfnamefont{J.}~\bibnamefont{Zhang}}, \bibinfo {author}
  {\bibfnamefont{I.~M.}\ \bibnamefont{Ross}},\ and\ \bibinfo {author}
  {\bibfnamefont{A.~G.}\ \bibnamefont{Cullis}},\ }%
  \bibfield{journal}{%
  \Doi{10.1103/PhysRevB.75.045338}{\bibinfo {journal} {Phys. Rev. B}}\ }%
  \textbf{\bibinfo {volume} {75}},\ \bibinfo {pages} {045338} (\bibinfo {year}
  {2007})%
  \bibAnnoteFile{NoStop}{PRBCalifano2007}%
\bibitem{NatureGraham2007}%
  \BibitemOpen
  \bibfield{author}{%
  \bibinfo {author} {\bibfnamefont{D.}~\bibnamefont{Graham-Rowe}},\ }%
  \bibfield{journal}{%
  \Doi{10.1038/nphoton.2006.85}{\bibinfo {journal} {Nat. Photonics}}\ }%
  \textbf{\bibinfo {volume} {1}},\ \bibinfo {pages} {75} (\bibinfo {year}
  {2007})%
  \bibAnnoteFile{NoStop}{NatureGraham2007}%
\bibitem{APLKumar2009}%
  \BibitemOpen
  \bibfield{author}{%
  \bibinfo {author} {\bibfnamefont{S.}~\bibnamefont{Kumar}}, \bibinfo {author}
  {\bibfnamefont{Q.}~\bibnamefont{Hu}},\ and\ \bibinfo {author}
  {\bibfnamefont{J.~L.}\ \bibnamefont{Reno}},\ }%
  \bibfield{journal}{%
  \Doi{10.1063/1.3114418}{\bibinfo {journal} {Appl. Phys. Lett.}}\ }%
  \textbf{\bibinfo {volume} {94}},\ \bibinfo {pages} {131105} (\bibinfo {year}
  {2009})%
  \bibAnnoteFile{NoStop}{APLKumar2009}%
\bibitem{APLAjili2005}%
  \BibitemOpen
  \bibfield{author}{%
  \bibinfo {author} {\bibfnamefont{L.}~\bibnamefont{Ajili}}, \bibinfo {author}
  {\bibfnamefont{G.}~\bibnamefont{Scalari}}, \bibinfo {author}
  {\bibfnamefont{N.}~\bibnamefont{Hoyler}}, \bibinfo {author}
  {\bibfnamefont{M.}~\bibnamefont{Giovannini}},\ and\ \bibinfo {author}
  {\bibfnamefont{J.}~\bibnamefont{Faist}},\ }%
  \bibfield{journal}{%
  \Doi{10.1063/1.2081122}{\bibinfo {journal} {Appl. Phys. Lett.}}\ }%
  \textbf{\bibinfo {volume} {87}},\ \bibinfo {pages} {141107} (\bibinfo {year}
  {2005})%
  \bibAnnoteFile{NoStop}{APLAjili2005}%
\bibitem{APLScalari2005}%
  \BibitemOpen
  \bibfield{author}{%
  \bibinfo {author} {\bibfnamefont{G.}~\bibnamefont{Scalari}}, \bibinfo
  {author} {\bibfnamefont{N.}~\bibnamefont{Hoyler}}, \bibinfo {author}
  {\bibfnamefont{M.}~\bibnamefont{Giovannini}},\ and\ \bibinfo {author}
  {\bibfnamefont{J.}~\bibnamefont{Faist}},\ }%
  \bibfield{journal}{%
  \Doi{10.1063/1.1920407}{\bibinfo {journal} {Appl. Phys. Lett.}}\ }%
  \textbf{\bibinfo {volume} {86}},\ \bibinfo {pages} {181101} (\bibinfo {year}
  {2005})%
  \bibAnnoteFile{NoStop}{APLScalari2005}%
\bibitem{APLLee2006}%
  \BibitemOpen
  \bibfield{author}{%
  \bibinfo {author} {\bibfnamefont{A.~W.~M.}\ \bibnamefont{Lee}}, \bibinfo
  {author} {\bibfnamefont{Q.}~\bibnamefont{Qin}}, \bibinfo {author}
  {\bibfnamefont{S.}~\bibnamefont{Kumar}}, \bibinfo {author}
  {\bibfnamefont{B.~S.}\ \bibnamefont{Williams}}, \bibinfo {author}
  {\bibfnamefont{Q.}~\bibnamefont{Hu}},\ and\ \bibinfo {author}
  {\bibfnamefont{J.~L.}\ \bibnamefont{Reno}},\ }%
  \bibfield{journal}{%
  \Doi{10.1063/1.2360210}{\bibinfo {journal} {Appl. Phys. Lett.}}\ }%
  \textbf{\bibinfo {volume} {89}},\ \bibinfo {pages} {141125} (\bibinfo {year}
  {2006})%
  \bibAnnoteFile{NoStop}{APLLee2006}%
\bibitem{ScienceDehlinger2000}%
  \BibitemOpen
  \bibfield{author}{%
  \bibinfo {author} {\bibfnamefont{G.}~\bibnamefont{Dehlinger}}, \bibinfo
  {author} {\bibfnamefont{L.}~\bibnamefont{Diehl}}, \bibinfo {author}
  {\bibfnamefont{U.}~\bibnamefont{Gennser}}, \bibinfo {author}
  {\bibfnamefont{H.}~\bibnamefont{Sigg}}, \bibinfo {author}
  {\bibfnamefont{J.}~\bibnamefont{Faist}}, \bibinfo {author}
  {\bibfnamefont{K.}~\bibnamefont{Ensslin}}, \bibinfo {author}
  {\bibfnamefont{D.}~\bibnamefont{Gr\"utzmacher}},\ and\ \bibinfo {author}
  {\bibfnamefont{E.}~\bibnamefont{M\"uller}},\ }%
  \bibfield{journal}{%
  \Doi{10.1126/science.290.5500.2277}{\bibinfo {journal} {Science}}\ }%
  \textbf{\bibinfo {volume} {290}},\ \bibinfo {pages} {2277} (\bibinfo {year}
  {2000})%
  \bibAnnoteFile{NoStop}{ScienceDehlinger2000}%
\bibitem{APLLynch2002}%
  \BibitemOpen
  \bibfield{author}{%
  \bibinfo {author} {\bibfnamefont{S.~A.}\ \bibnamefont{Lynch}}, \bibinfo
  {author} {\bibfnamefont{R.}~\bibnamefont{Bates}}, \bibinfo {author}
  {\bibfnamefont{D.~J.}\ \bibnamefont{Paul}}, \bibinfo {author}
  {\bibfnamefont{D.~J.}\ \bibnamefont{Norris}}, \bibinfo {author}
  {\bibfnamefont{A.~G.}\ \bibnamefont{Cullis}}, \bibinfo {author}
  {\bibfnamefont{Z.}~\bibnamefont{Ikoni\'c}}, \bibinfo {author}
  {\bibfnamefont{R.~W.}\ \bibnamefont{Kelsall}}, \bibinfo {author}
  {\bibfnamefont{P.}~\bibnamefont{Harrison}}, \bibinfo {author}
  {\bibfnamefont{D.~D.}\ \bibnamefont{Arnone}},\ and\ \bibinfo {author}
  {\bibfnamefont{C.~R.}\ \bibnamefont{Pidgeon}},\ }%
  \bibfield{journal}{%
  \Doi{10.1063/1.1501759}{\bibinfo {journal} {Appl. Phys. Lett.}}\ }%
  \textbf{\bibinfo {volume} {81}},\ \bibinfo {pages} {1543} (\bibinfo {year}
  {2002})%
  \bibAnnoteFile{NoStop}{APLLynch2002}%
\bibitem{IJHSESKelsall2003}%
  \BibitemOpen
  \bibfield{author}{%
  \bibinfo {author} {\bibfnamefont{R.~W.}\ \bibnamefont{Kelsall}}\ and\
  \bibinfo {author} {\bibfnamefont{R.~A.}\ \bibnamefont{Soref}},\ }%
  \bibfield{journal}{%
  \Doi{10.1142/S012915640300182X}{\bibinfo {journal} {Int. J. High Speed
  Electron.}}\ }%
  \textbf{\bibinfo {volume} {13}},\ \bibinfo {pages} {547} (\bibinfo {year}
  {2003})%
  \bibAnnoteFile{NoStop}{IJHSESKelsall2003}%
\bibitem{APLLever2009}%
  \BibitemOpen
  \bibfield{author}{%
  \bibinfo {author} {\bibfnamefont{L.}~\bibnamefont{Lever}}, \bibinfo {author}
  {\bibfnamefont{A.}~\bibnamefont{Valavanis}}, \bibinfo {author}
  {\bibfnamefont{C.~A.}\ \bibnamefont{Evans}}, \bibinfo {author}
  {\bibfnamefont{Z.}~\bibnamefont{Ikoni\'{c}}},\ and\ \bibinfo {author}
  {\bibfnamefont{R.~W.}\ \bibnamefont{Kelsall}},\ }%
  \bibfield{journal}{%
  \Doi{10.1063/1.3237177}{\bibinfo {journal} {Appl. Phys. Lett.}}\ }%
  \textbf{\bibinfo {volume} {95}},\ \bibinfo {pages} {131103} (\bibinfo {year}
  {2009})%
  \bibAnnoteFile{NoStop}{APLLever2009}%
\bibitem{APLDriscoll2006}%
  \BibitemOpen
  \bibfield{author}{%
  \bibinfo {author} {\bibfnamefont{K.}~\bibnamefont{Driscoll}}\ and\ \bibinfo
  {author} {\bibfnamefont{R.}~\bibnamefont{Paiella}},\ }%
  \bibfield{journal}{%
  \Doi{10.1063/1.2385861}{\bibinfo {journal} {Appl. Phys. Lett.}}\ }%
  \textbf{\bibinfo {volume} {89}},\ \bibinfo {pages} {191110} (\bibinfo {year}
  {2006})%
  \bibAnnoteFile{NoStop}{APLDriscoll2006}%
\bibitem{JAPDriscoll2007}%
  \BibitemOpen
  \bibfield{author}{%
  \bibinfo {author} {\bibfnamefont{K.}~\bibnamefont{Driscoll}}\ and\ \bibinfo
  {author} {\bibfnamefont{R.}~\bibnamefont{Paiella}},\ }%
  \bibfield{journal}{%
  \Doi{10.1063/1.2803896}{\bibinfo {journal} {J. Appl. Phys.}}\ }%
  \textbf{\bibinfo {volume} {102}},\ \bibinfo {pages} {093103} (\bibinfo {year}
  {2007})%
  \bibAnnoteFile{NoStop}{JAPDriscoll2007}%
\bibitem{PRBValavanis2008_2}%
  \BibitemOpen
  \bibfield{author}{%
  \bibinfo {author} {\bibfnamefont{A.}~\bibnamefont{Valavanis}}, \bibinfo
  {author} {\bibfnamefont{L.}~\bibnamefont{Lever}}, \bibinfo {author}
  {\bibfnamefont{C.~A.}\ \bibnamefont{Evans}}, \bibinfo {author}
  {\bibfnamefont{Z.}~\bibnamefont{Ikoni\'{c}}},\ and\ \bibinfo {author}
  {\bibfnamefont{R.~W.}\ \bibnamefont{Kelsall}},\ }%
  \bibfield{journal}{%
  \Doi{10.1103/PhysRevB.78.035420}{\bibinfo {journal} {Phys. Rev. B}}\ }%
  \textbf{\bibinfo {volume} {78}},\ \bibinfo {pages} {035420} (\bibinfo {year}
  {2008})%
  \bibAnnoteFile{NoStop}{PRBValavanis2008_2}%
\bibitem{APLLever2008}%
  \BibitemOpen
  \bibfield{author}{%
  \bibinfo {author} {\bibfnamefont{L.}~\bibnamefont{Lever}}, \bibinfo {author}
  {\bibfnamefont{A.}~\bibnamefont{Valavanis}}, \bibinfo {author}
  {\bibfnamefont{Z.}~\bibnamefont{Ikoni\'c}},\ and\ \bibinfo {author}
  {\bibfnamefont{R.~W.}\ \bibnamefont{Kelsall}},\ }%
  \bibfield{journal}{%
  \Doi{10.1063/1.2836023}{\bibinfo {journal} {Appl. Phys. Lett.}}\ }%
  \textbf{\bibinfo {volume} {92}},\ \bibinfo {pages} {021124} (\bibinfo {year}
  {2008})%
  \bibAnnoteFile{NoStop}{APLLever2008}%
\bibitem{SSTHan2007}%
  \BibitemOpen
  \bibfield{author}{%
  \bibinfo {author} {\bibfnamefont{G.}~\bibnamefont{Han}}\ and\ \bibinfo
  {author} {\bibfnamefont{J.}~\bibnamefont{Yu}},\ }%
  \bibfield{journal}{%
  \Doi{10.1088/0268-1242/22/7/016}{\bibinfo {journal} {Semicond. Sci.
  Technol.}}\ }%
  \textbf{\bibinfo {volume} {22}},\ \bibinfo {pages} {769} (\bibinfo {year}
  {2007})%
  \bibAnnoteFile{NoStop}{SSTHan2007}%
\bibitem{APLSun2007}%
  \BibitemOpen
  \bibfield{author}{%
  \bibinfo {author} {\bibfnamefont{G.}~\bibnamefont{Sun}}, \bibinfo {author}
  {\bibfnamefont{H.~H.}\ \bibnamefont{Cheng}}, \bibinfo {author}
  {\bibfnamefont{J.}~\bibnamefont{Men\'{e}ndez}}, \bibinfo {author}
  {\bibfnamefont{J.~B.}\ \bibnamefont{Khurgin}},\ and\ \bibinfo {author}
  {\bibfnamefont{R.~A.}\ \bibnamefont{Soref}},\ }%
  \bibfield{journal}{%
  \Doi{10.1063/1.2749844}{\bibinfo {journal} {Appl. Phys. Lett.}}\ }%
  \textbf{\bibinfo {volume} {90}},\ \bibinfo {pages} {251105} (\bibinfo {year}
  {2007})%
  \bibAnnoteFile{NoStop}{APLSun2007}%
\bibitem{JPhysChemDismukes1964}%
  \BibitemOpen
  \bibfield{author}{%
  \bibinfo {author} {\bibfnamefont{J.~P.}\ \bibnamefont{Dismukes}}, \bibinfo
  {author} {\bibfnamefont{L.}~\bibnamefont{Ekstrom}},\ and\ \bibinfo {author}
  {\bibfnamefont{R.~J.}\ \bibnamefont{Paff}},\ }%
  \bibfield{journal}{%
  \Doi{10.1021/j100792a049}{\bibinfo {journal} {J. Phys. Chem.}}\ }%
  \textbf{\bibinfo {volume} {68}},\ \bibinfo {pages} {3021} (\bibinfo {year}
  {1964})%
  \bibAnnoteFile{NoStop}{JPhysChemDismukes1964}%
\bibitem{PRLZwerdling1960}%
  \BibitemOpen
  \bibfield{author}{%
  \bibinfo {author} {\bibfnamefont{S.}~\bibnamefont{Zwerdling}}, \bibinfo
  {author} {\bibfnamefont{K.~J.}\ \bibnamefont{Button}}, \bibinfo {author}
  {\bibfnamefont{B.}~\bibnamefont{Lax}},\ and\ \bibinfo {author}
  {\bibfnamefont{L.~M.}\ \bibnamefont{Roth}},\ }%
  \bibfield{journal}{%
  \Doi{10.1103/PhysRevLett.4.173}{\bibinfo {journal} {Phys. Rev. Lett.}}\ }%
  \textbf{\bibinfo {volume} {4}},\ \bibinfo {pages} {173} (\bibinfo {year}
  {1960})%
  \bibAnnoteFile{NoStop}{PRLZwerdling1960}%
\bibitem{JPCSKane1956}%
  \BibitemOpen
  \bibfield{author}{%
  \bibinfo {author} {\bibfnamefont{E.~O.}\ \bibnamefont{Kane}},\ }%
  \bibfield{journal}{%
  \Doi{10.1016/0022-3697(56)90014-2}{\bibinfo {journal} {J. Phys. Chem.
  Solids}}\ }%
  \textbf{\bibinfo {volume} {1}},\ \bibinfo {pages} {82} (\bibinfo {year}
  {1956})%
  \bibAnnoteFile{NoStop}{JPCSKane1956}%
\bibitem{JAPMcSkimin1964}%
  \BibitemOpen
  \bibfield{author}{%
  \bibinfo {author} {\bibfnamefont{H.~J.}\ \bibnamefont{McSkimin}}\ and\
  \bibinfo {author} {\bibnamefont{{P. Andreatch Jr.}}},\ }%
  \bibfield{journal}{%
  \Doi{10.1063/1.1713214}{\bibinfo {journal} {J. Appl. Phys.}}\ }%
  \textbf{\bibinfo {volume} {35}},\ \bibinfo {pages} {3312} (\bibinfo {year}
  {1964})%
  \bibAnnoteFile{NoStop}{JAPMcSkimin1964}%
\bibitem{PRBVanDeWalle1986}%
  \BibitemOpen
  \bibfield{author}{%
  \bibinfo {author} {\bibfnamefont{C.~G.}\ \bibnamefont{{Van de Walle}}}\ and\
  \bibinfo {author} {\bibfnamefont{R.~M.}\ \bibnamefont{Martin}},\ }%
  \bibfield{journal}{%
  \Doi{10.1103/PhysRevB.34.5621}{\bibinfo {journal} {Phys. Rev. B}}\ }%
  \textbf{\bibinfo {volume} {34}},\ \bibinfo {pages} {5621} (\bibinfo {year}
  {1986})%
  \bibAnnoteFile{NoStop}{PRBVanDeWalle1986}%
\bibitem{PRBRieger1993}%
  \BibitemOpen
  \bibfield{author}{%
  \bibinfo {author} {\bibfnamefont{M.~M.}\ \bibnamefont{Rieger}}\ and\ \bibinfo
  {author} {\bibfnamefont{P.}~\bibnamefont{Vogl}},\ }%
  \bibfield{journal}{%
  \Doi{10.1103/PhysRevB.48.14276}{\bibinfo {journal} {Phys. Rev. B}}\ }%
  \textbf{\bibinfo {volume} {48}},\ \bibinfo {pages} {14276} (\bibinfo {year}
  {1993})%
  \bibAnnoteFile{NoStop}{PRBRieger1993}%
\bibitem{APLAfanas'ev2009}%
  \BibitemOpen
  \bibfield{author}{%
  \bibinfo {author} {\bibfnamefont{V.~V.}\ \bibnamefont{Afanas'ev}}, \bibinfo
  {author} {\bibfnamefont{A.}~\bibnamefont{Stesmans}}, \bibinfo {author}
  {\bibfnamefont{L.}~\bibnamefont{Souriau}}, \bibinfo {author}
  {\bibfnamefont{R.}~\bibnamefont{Loo}},\ and\ \bibinfo {author}
  {\bibfnamefont{M.}~\bibnamefont{Meuris}},\ }%
  \bibfield{journal}{%
  \Doi{10.1063/1.3125434}{\bibinfo {journal} {Appl. Phys. Lett.}}\ }%
  \textbf{\bibinfo {volume} {94}},\ \bibinfo {pages} {172106} (\bibinfo {year}
  {2009})%
  \bibAnnoteFile{NoStop}{APLAfanas'ev2009}%
\bibitem{PRBBusby2010}%
  \BibitemOpen
  \bibfield{author}{%
  \bibinfo {author} {\bibfnamefont{Y.}~\bibnamefont{Busby}}, \bibinfo {author}
  {\bibfnamefont{M.}~\bibnamefont{De~Seta}}, \bibinfo {author}
  {\bibfnamefont{G.}~\bibnamefont{Capellini}}, \bibinfo {author}
  {\bibfnamefont{F.}~\bibnamefont{Evangelisti}}, \bibinfo {author}
  {\bibfnamefont{M.}~\bibnamefont{Ortolani}}, \bibinfo {author}
  {\bibfnamefont{M.}~\bibnamefont{Virgilio}}, \bibinfo {author}
  {\bibfnamefont{G.}~\bibnamefont{Grosso}}, \bibinfo {author}
  {\bibfnamefont{G.}~\bibnamefont{Pizzi}}, \bibinfo {author}
  {\bibfnamefont{P.}~\bibnamefont{Calvani}}, \bibinfo {author}
  {\bibfnamefont{S.}~\bibnamefont{Lupi}}, \bibinfo {author}
  {\bibfnamefont{M.}~\bibnamefont{Nardone}}, \bibinfo {author}
  {\bibfnamefont{G.}~\bibnamefont{Nicotra}},\ and\ \bibinfo {author}
  {\bibfnamefont{C.}~\bibnamefont{Spinella}},\ }%
  \bibfield{journal}{%
  \Doi{10.1103/PhysRevB.82.205317}{\bibinfo {journal} {Phys. Rev. B}}\ }%
  \textbf{\bibinfo {volume} {82}},\ \bibinfo {pages} {205317} (\bibinfo {year}
  {2010})%
  \bibAnnoteFile{NoStop}{PRBBusby2010}%
\bibitem{PRBWeber1989}%
  \BibitemOpen
  \bibfield{author}{%
  \bibinfo {author} {\bibfnamefont{J.}~\bibnamefont{Weber}}\ and\ \bibinfo
  {author} {\bibfnamefont{M.~I.}\ \bibnamefont{Alonso}},\ }%
  \bibfield{journal}{%
  \Doi{10.1103/PhysRevB.40.5683}{\bibinfo {journal} {Phys. Rev. B}}\ }%
  \textbf{\bibinfo {volume} {40}},\ \bibinfo {pages} {5683} (\bibinfo {year}
  {1989})%
  \bibAnnoteFile{NoStop}{PRBWeber1989}%
\bibitem{SSTPaul2004}%
  \BibitemOpen
  \bibfield{author}{%
  \bibinfo {author} {\bibfnamefont{D.~J.}\ \bibnamefont{Paul}},\ }%
  \bibfield{journal}{%
  \Doi{10.1088/0268-1242/19/10/R02}{\bibinfo {journal} {Semicond. Sci.
  Technol.}}\ }%
  \textbf{\bibinfo {volume} {19}},\ \bibinfo {pages} {R75} (\bibinfo {year}
  {2004})%
  \bibAnnoteFile{NoStop}{SSTPaul2004}%
\bibitem{SSESmirnov2004}%
  \BibitemOpen
  \bibfield{author}{%
  \bibinfo {author} {\bibfnamefont{S.}~\bibnamefont{Smirnov}}\ and\ \bibinfo
  {author} {\bibfnamefont{H.}~\bibnamefont{Kosina}},\ }%
  \bibfield{journal}{%
  \Doi{doi:10.1016/j.sse.2004.01.014}{\bibinfo {journal} {Sol. State
  Electron.}}\ }%
  \textbf{\bibinfo {volume} {48}},\ \bibinfo {pages} {1325} (\bibinfo {year}
  {2004})%
  \bibAnnoteFile{NoStop}{SSESmirnov2004}%
\bibitem{JAPRahman2005}%
  \BibitemOpen
  \bibfield{author}{%
  \bibinfo {author} {\bibfnamefont{A.}~\bibnamefont{Rahman}}, \bibinfo {author}
  {\bibfnamefont{M.~S.}\ \bibnamefont{Lundstrom}},\ and\ \bibinfo {author}
  {\bibfnamefont{A.~W.}\ \bibnamefont{Ghosh}},\ }%
  \bibfield{journal}{%
  \Doi{10.1063/1.1845586}{\bibinfo {journal} {J. Appl. Phys.}}\ }%
  \textbf{\bibinfo {volume} {97}},\ \bibinfo {pages} {053702} (\bibinfo {year}
  {2005})%
  \bibAnnoteFile{NoStop}{JAPRahman2005}%
\bibitem{NanotechCuratola2002}%
  \BibitemOpen
  \bibfield{author}{%
  \bibinfo {author} {\bibfnamefont{G.}~\bibnamefont{Curatola}}\ and\ \bibinfo
  {author} {\bibfnamefont{G.}~\bibnamefont{Iannaccone}},\ }%
  \bibfield{journal}{%
  \Doi{10.1088/0957-4484/13/3/306}{\bibinfo {journal} {Nanotechnology}}\ }%
  \textbf{\bibinfo {volume} {13}},\ \bibinfo {pages} {267} (\bibinfo {year}
  {2002})%
  \bibAnnoteFile{NoStop}{NanotechCuratola2002}%
\bibitem{JAPRamMohan2004}%
  \BibitemOpen
  \bibfield{author}{%
  \bibinfo {author} {\bibfnamefont{L.~R.}\ \bibnamefont{Ram-Mohan}}, \bibinfo
  {author} {\bibfnamefont{K.~H.}\ \bibnamefont{Yoo}},\ and\ \bibinfo {author}
  {\bibfnamefont{J.}~\bibnamefont{Moussa}},\ }%
  \bibfield{journal}{%
  \Doi{10.1063/1.1649458}{\bibinfo {journal} {J. Appl. Phys.}}\ }%
  \textbf{\bibinfo {volume} {95}},\ \bibinfo {pages} {3081} (\bibinfo {year}
  {2004})%
  \bibAnnoteFile{NoStop}{JAPRamMohan2004}%
\bibitem{PRBVirgilio2009}%
  \BibitemOpen
  \bibfield{author}{%
  \bibinfo {author} {\bibfnamefont{M.}~\bibnamefont{Virgilio}}\ and\ \bibinfo
  {author} {\bibfnamefont{G.}~\bibnamefont{Grosso}},\ }%
  \bibfield{journal}{%
  \Doi{10.1103/PhysRevB.79.165310}{\bibinfo {journal} {Phys. Rev. B}}\ }%
  \textbf{\bibinfo {volume} {79}},\ \bibinfo {pages} {165310} (\bibinfo {year}
  {2009})%
  \bibAnnoteFile{NoStop}{PRBVirgilio2009}%
\bibitem{PRBValavanis2007}%
  \BibitemOpen
  \bibfield{author}{%
  \bibinfo {author} {\bibfnamefont{A.}~\bibnamefont{Valavanis}}, \bibinfo
  {author} {\bibfnamefont{Z.}~\bibnamefont{Ikoni\ifmmode~\acute{c}\else
  \'{c}\fi{}}},\ and\ \bibinfo {author} {\bibfnamefont{R.~W.}\
  \bibnamefont{Kelsall}},\ }%
  \bibfield{journal}{%
  \Doi{10.1103/PhysRevB.75.205332}{\bibinfo {journal} {Phys. Rev. B}}\ }%
  \textbf{\bibinfo {volume} {75}},\ \bibinfo {pages} {205332} (\bibinfo {year}
  {2007})%
  \bibAnnoteFile{NoStop}{PRBValavanis2007}%
\bibitem{PRBValavanis2008}%
  \BibitemOpen
  \bibfield{author}{%
  \bibinfo {author} {\bibfnamefont{A.}~\bibnamefont{Valavanis}}, \bibinfo
  {author} {\bibfnamefont{Z.}~\bibnamefont{Ikoni\'c}},\ and\ \bibinfo {author}
  {\bibfnamefont{R.~W.}\ \bibnamefont{Kelsall}},\ }%
  \bibfield{journal}{%
  \Doi{10.1103/PhysRevB.77.075312}{\bibinfo {journal} {Phys. Rev. B}}\ }%
  \textbf{\bibinfo {volume} {77}},\ \bibinfo {pages} {075312} (\bibinfo {year}
  {2008})%
  \bibAnnoteFile{NoStop}{PRBValavanis2008}%
\bibitem{JAPJovanovic2006}%
  \BibitemOpen
  \bibfield{author}{%
  \bibinfo {author} {\bibfnamefont{V.~D.}\ \bibnamefont{Jovanovi\'c}}, \bibinfo
  {author} {\bibfnamefont{S.}~\bibnamefont{H\"ofling}}, \bibinfo {author}
  {\bibfnamefont{D.}~\bibnamefont{Indjin}}, \bibinfo {author}
  {\bibfnamefont{N.}~\bibnamefont{Vukmirovi\'c}}, \bibinfo {author}
  {\bibfnamefont{Z.}~\bibnamefont{Ikoni\'c}}, \bibinfo {author}
  {\bibfnamefont{P.}~\bibnamefont{Harrison}}, \bibinfo {author}
  {\bibfnamefont{J.~P.}\ \bibnamefont{Reithmaier}},\ and\ \bibinfo {author}
  {\bibfnamefont{A.}~\bibnamefont{Forchel}},\ }%
  \bibfield{journal}{%
  \Doi{10.1063/1.2194312}{\bibinfo {journal} {J. Appl. Phys.}}\ }%
  \textbf{\bibinfo {volume} {99}},\ \bibinfo {pages} {103106} (\bibinfo {year}
  {2006})%
  \bibAnnoteFile{NoStop}{JAPJovanovic2006}%
\bibitem{SSTIndjin2005}%
  \BibitemOpen
  \bibfield{author}{%
  \bibinfo {author} {\bibfnamefont{D.}~\bibnamefont{Indjin}}, \bibinfo {author}
  {\bibfnamefont{Z.}~\bibnamefont{Ikoni\'c}}, \bibinfo {author}
  {\bibfnamefont{V.~D.}\ \bibnamefont{Jovanovi\'c}}, \bibinfo {author}
  {\bibfnamefont{N.}~\bibnamefont{Vukmirovi\'c}}, \bibinfo {author}
  {\bibfnamefont{P.}~\bibnamefont{Harrison}},\ and\ \bibinfo {author}
  {\bibfnamefont{R.~W.}\ \bibnamefont{Kelsall}},\ }%
  \bibfield{journal}{%
  \Doi{10.1088/0268-1242/20/7/014}{\bibinfo {journal} {Semicond. Sci.
  Technol.}}\ }%
  \textbf{\bibinfo {volume} {20}},\ \bibinfo {pages} {S237} (\bibinfo {year}
  {2005})%
  \bibAnnoteFile{NoStop}{SSTIndjin2005}%
\bibitem{PRBLee2002}%
  \BibitemOpen
  \bibfield{author}{%
  \bibinfo {author} {\bibfnamefont{S.-C.}\ \bibnamefont{Lee}}\ and\ \bibinfo
  {author} {\bibfnamefont{A.}~\bibnamefont{Wacker}},\ }%
  \bibfield{journal}{%
  \Doi{10.1103/PhysRevB.66.245314}{\bibinfo {journal} {Phys. Rev. B}}\ }%
  \textbf{\bibinfo {volume} {66}},\ \bibinfo {pages} {245314} (\bibinfo {year}
  {2002})%
  \bibAnnoteFile{NoStop}{PRBLee2002}%
\bibitem{PRBWang2010}%
  \BibitemOpen
  \bibfield{author}{%
  \bibinfo {author} {\bibfnamefont{F.}~\bibnamefont{Wang}}, \bibinfo {author}
  {\bibfnamefont{X.~G.}\ \bibnamefont{Guo}},\ and\ \bibinfo {author}
  {\bibfnamefont{J.~C.}\ \bibnamefont{Cao}},\ }%
  \bibfield{journal}{%
  \Doi{10.1103/PhysRevB.81.045308}{\bibinfo {journal} {Phys. Rev. B}}\ }%
  \textbf{\bibinfo {volume} {81}},\ \bibinfo {pages} {045308} (\bibinfo {year}
  {2010})%
  \bibAnnoteFile{NoStop}{PRBWang2010}%
\bibitem{LPRevScalari2009}%
  \BibitemOpen
  \bibfield{author}{%
  \bibinfo {author} {\bibfnamefont{G.}~\bibnamefont{Scalari}}, \bibinfo
  {author} {\bibfnamefont{C.}~\bibnamefont{Walther}}, \bibinfo {author}
  {\bibfnamefont{M.}~\bibnamefont{Fischer}}, \bibinfo {author}
  {\bibfnamefont{R.}~\bibnamefont{Terazzi}}, \bibinfo {author}
  {\bibfnamefont{H.}~\bibnamefont{Beere}}, \bibinfo {author}
  {\bibfnamefont{D.}~\bibnamefont{Ritchie}},\ and\ \bibinfo {author}
  {\bibfnamefont{J.}~\bibnamefont{Faist}},\ }%
  \bibfield{journal}{%
  \Doi{10.1002/lpor.200810030}{\bibinfo {journal} {Laser and Photon. Rev.}}\ }%
  \textbf{\bibinfo {volume} {3}},\ \bibinfo {pages} {45} (\bibinfo {year}
  {2009})%
  \bibAnnoteFile{NoStop}{LPRevScalari2009}%
\bibitem{JAPCallebaut2005}%
  \BibitemOpen
  \bibfield{author}{%
  \bibinfo {author} {\bibfnamefont{H.}~\bibnamefont{Callebaut}}\ and\ \bibinfo
  {author} {\bibfnamefont{Q.}~\bibnamefont{Hu}},\ }%
  \bibfield{journal}{%
  \Doi{10.1063/1.2136420}{\bibinfo {journal} {J. Appl. Phys.}}\ }%
  \textbf{\bibinfo {volume} {98}},\ \bibinfo {pages} {104505} (\bibinfo {year}
  {2005})%
  \bibAnnoteFile{NoStop}{JAPCallebaut2005}%
\bibitem{PRBDupont2010}%
  \BibitemOpen
  \bibfield{author}{%
  \bibinfo {author} {\bibfnamefont{E.}~\bibnamefont{Dupont}}, \bibinfo {author}
  {\bibfnamefont{S.}~\bibnamefont{Fathololoumi}},\ and\ \bibinfo {author}
  {\bibfnamefont{H.~C.}\ \bibnamefont{Liu}},\ }%
  \bibfield{journal}{%
  \Doi{10.1103/PhysRevB.81.205311}{\bibinfo {journal} {Phys. Rev. B}}\ }%
  \textbf{\bibinfo {volume} {81}},\ \bibinfo {pages} {205311} (\bibinfo {year}
  {2010})%
  \bibAnnoteFile{NoStop}{PRBDupont2010}%
\bibitem{PRBKumar2009}%
  \BibitemOpen
  \bibfield{author}{%
  \bibinfo {author} {\bibfnamefont{S.}~\bibnamefont{Kumar}}\ and\ \bibinfo
  {author} {\bibfnamefont{Q.}~\bibnamefont{Hu}},\ }%
  \bibfield{journal}{%
  \Doi{10.1103/PhysRevB.80.245316}{\bibinfo {journal} {Phys. Rev. B}}\ }%
  \textbf{\bibinfo {volume} {80}},\ \bibinfo {pages} {245316} (\bibinfo {year}
  {2009})%
  \bibAnnoteFile{NoStop}{PRBKumar2009}%
\bibitem{PRBGordon2009}%
  \BibitemOpen
  \bibfield{author}{%
  \bibinfo {author} {\bibfnamefont{A.}~\bibnamefont{Gordon}}\ and\ \bibinfo
  {author} {\bibfnamefont{D.}~\bibnamefont{Majer}},\ }%
  \bibfield{journal}{%
  \Doi{10.1103/PhysRevB.80.195317}{\bibinfo {journal} {Phys. Rev. B}}\ }%
  \textbf{\bibinfo {volume} {80}},\ \bibinfo {pages} {195317} (\bibinfo {year}
  {2009})%
  \bibAnnoteFile{NoStop}{PRBGordon2009}%
\bibitem{PRBQuang2007}%
  \BibitemOpen
  \bibfield{author}{%
  \bibinfo {author} {\bibfnamefont{D.~N.}\ \bibnamefont{Quang}}, \bibinfo
  {author} {\bibfnamefont{N.~H.}\ \bibnamefont{Tung}}, \bibinfo {author}
  {\bibfnamefont{D.~T.}\ \bibnamefont{Hien}},\ and\ \bibinfo {author}
  {\bibfnamefont{H.~A.}\ \bibnamefont{Huy}},\ }%
  \bibfield{journal}{%
  \Doi{10.1103/PhysRevB.75.073305}{\bibinfo {journal} {Phys. Rev. B}}\ }%
  \textbf{\bibinfo {volume} {75}},\ \bibinfo {pages} {073305} (\bibinfo {year}
  {2007})%
  \bibAnnoteFile{NoStop}{PRBQuang2007}%
\bibitem{PRLMurphy-Armando2006}%
  \BibitemOpen
  \bibfield{author}{%
  \bibinfo {author} {\bibfnamefont{F.}~\bibnamefont{Murphy-Armando}}\ and\
  \bibinfo {author} {\bibfnamefont{S.}~\bibnamefont{Fahy}},\ }%
  \bibfield{journal}{%
  \Doi{10.1103/PhysRevLett.97.096606}{\bibinfo {journal} {Phys. Rev. Lett.}}\
  }%
  \textbf{\bibinfo {volume} {97}},\ \bibinfo {pages} {096606} (\bibinfo {year}
  {2006})%
  \bibAnnoteFile{NoStop}{PRLMurphy-Armando2006}%
\bibitem{JAPUnuma2003}%
  \BibitemOpen
  \bibfield{author}{%
  \bibinfo {author} {\bibfnamefont{T.}~\bibnamefont{Unuma}}, \bibinfo {author}
  {\bibfnamefont{M.}~\bibnamefont{Yoshita}}, \bibinfo {author}
  {\bibfnamefont{T.}~\bibnamefont{Noda}}, \bibinfo {author}
  {\bibfnamefont{H.}~\bibnamefont{Sakaki}},\ and\ \bibinfo {author}
  {\bibfnamefont{H.}~\bibnamefont{Akiyama}},\ }%
  \bibfield{journal}{%
  \Doi{10.1063/1.1535733}{\bibinfo {journal} {J. Appl. Phys.}}\ }%
  \textbf{\bibinfo {volume} {93}},\ \bibinfo {pages} {1586} (\bibinfo {year}
  {2003})%
  \bibAnnoteFile{NoStop}{JAPUnuma2003}%
\bibitem{JAPSmet1996}%
  \BibitemOpen
  \bibfield{author}{%
  \bibinfo {author} {\bibfnamefont{J.~H.}\ \bibnamefont{Smet}}, \bibinfo
  {author} {\bibfnamefont{C.~G.}\ \bibnamefont{Fonstad}},\ and\ \bibinfo
  {author} {\bibfnamefont{Q.}~\bibnamefont{Hu}},\ }%
  \bibfield{journal}{%
  \Doi{10.1063/1.362607}{\bibinfo {journal} {J. Appl. Phys.}}\ }%
  \textbf{\bibinfo {volume} {79}},\ \bibinfo {pages} {9305} (\bibinfo {year}
  {1996})%
  \bibAnnoteFile{NoStop}{JAPSmet1996}%
\bibitem{PRBFischetti1993}%
  \BibitemOpen
  \bibfield{author}{%
  \bibinfo {author} {\bibfnamefont{M.~V.}\ \bibnamefont{Fischetti}}\ and\
  \bibinfo {author} {\bibfnamefont{S.~E.}\ \bibnamefont{Laux}},\ }%
  \bibfield{journal}{%
  \Doi{10.1103/PhysRevB.48.2244}{\bibinfo {journal} {Phys. Rev. B}}\ }%
  \textbf{\bibinfo {volume} {48}},\ \bibinfo {pages} {2244} (\bibinfo {year}
  {1993})%
  \bibAnnoteFile{NoStop}{PRBFischetti1993}%
\bibitem{Ridley1997}%
  \BibitemOpen
  \bibfield{author}{%
  \bibinfo {author} {\bibfnamefont{B.~K.}\ \bibnamefont{Ridley}},\ }%
  \emph{\bibinfo {title} {Electrons and phonons in semiconductor multilayers}}\
  (\bibinfo {publisher} {Cambridge University Press, Cambridge},\ \bibinfo
  {year} {1997})%
  \bibAnnoteFile{NoStop}{Ridley1997}%
\bibitem{IEEEFischetti1991}%
  \BibitemOpen
  \bibfield{author}{%
  \bibinfo {author} {\bibfnamefont{M.}~\bibnamefont{Fischetti}},\ }%
  \bibfield{journal}{%
  \Doi{10.1109/16.75176}{\bibinfo {journal} {IEEE Trans. Electron. Dev.}}\ }%
  \textbf{\bibinfo {volume} {38}},\ \bibinfo {pages} {634} (\bibinfo {year}
  {1991})%
  \bibAnnoteFile{NoStop}{IEEEFischetti1991}%
\bibitem{PRBJacoboni1981}%
  \BibitemOpen
  \bibfield{author}{%
  \bibinfo {author} {\bibfnamefont{C.}~\bibnamefont{Jacoboni}}, \bibinfo
  {author} {\bibfnamefont{F.}~\bibnamefont{Nava}}, \bibinfo {author}
  {\bibfnamefont{C.}~\bibnamefont{Canali}},\ and\ \bibinfo {author}
  {\bibfnamefont{G.}~\bibnamefont{Ottaviani}},\ }%
  \bibfield{journal}{%
  \Doi{10.1103/PhysRevB.24.1014}{\bibinfo {journal} {Phys. Rev. B}}\ }%
  \textbf{\bibinfo {volume} {24}},\ \bibinfo {pages} {1014} (\bibinfo {year}
  {1981})%
  \bibAnnoteFile{NoStop}{PRBJacoboni1981}%
\bibitem{JAPDollfus1997}%
  \BibitemOpen
  \bibfield{author}{%
  \bibinfo {author} {\bibfnamefont{P.}~\bibnamefont{Dollfus}},\ }%
  \bibfield{journal}{%
  \Doi{10.1063/1.365696}{\bibinfo {journal} {J. Appl. Phys.}}\ }%
  \textbf{\bibinfo {volume} {82}},\ \bibinfo {pages} {3911} (\bibinfo {year}
  {1997})%
  \bibAnnoteFile{NoStop}{JAPDollfus1997}%
\bibitem{PRBFerry1976}%
  \BibitemOpen
  \bibfield{author}{%
  \bibinfo {author} {\bibfnamefont{D.~K.}\ \bibnamefont{Ferry}},\ }%
  \bibfield{journal}{%
  \Doi{10.1103/PhysRevB.14.1605}{\bibinfo {journal} {Phys. Rev. B}}\ }%
  \textbf{\bibinfo {volume} {14}},\ \bibinfo {pages} {1605} (\bibinfo {year}
  {1976})%
  \bibAnnoteFile{NoStop}{PRBFerry1976}%
\bibitem{PRBMonsef2002}%
  \BibitemOpen
  \bibfield{author}{%
  \bibinfo {author} {\bibfnamefont{F.}~\bibnamefont{Monsef}}, \bibinfo {author}
  {\bibfnamefont{P.}~\bibnamefont{Dollfus}}, \bibinfo {author}
  {\bibfnamefont{S.}~\bibnamefont{Galdin}},\ and\ \bibinfo {author}
  {\bibfnamefont{A.}~\bibnamefont{Bournel}},\ }%
  \bibfield{journal}{%
  \Doi{10.1103/PhysRevB.65.212304}{\bibinfo {journal} {Phys. Rev. B}}\ }%
  \textbf{\bibinfo {volume} {65}},\ \bibinfo {pages} {212304} (\bibinfo {year}
  {2002})%
  \bibAnnoteFile{NoStop}{PRBMonsef2002}%
\bibitem{PRBMonsef2002Errata}%
  \BibitemOpen
  \bibfield{author}{%
  \bibinfo {author} {\bibfnamefont{F.}~\bibnamefont{Monsef}}, \bibinfo {author}
  {\bibfnamefont{P.}~\bibnamefont{Dollfus}}, \bibinfo {author}
  {\bibfnamefont{S.}~\bibnamefont{Galdin}},\ and\ \bibinfo {author}
  {\bibfnamefont{A.}~\bibnamefont{Bournel}},\ }%
  \bibfield{journal}{%
  \Doi{10.1103/PhysRevB.67.059903}{\bibinfo {journal} {Phys. Rev. B}}\ }%
  \textbf{\bibinfo {volume} {67}},\ \bibinfo {pages} {059903(E)} (\bibinfo
  {year} {2003})%
  \bibAnnoteFile{NoStop}{PRBMonsef2002Errata}%
\bibitem{Davies1998}%
  \BibitemOpen
  \bibfield{author}{%
  \bibinfo {author} {\bibfnamefont{J.~H.}\ \bibnamefont{Davies}},\ }%
  \emph{\bibinfo {title} {The Physics of Low-Dimensional Semiconductors: An
  Introduction}}\ (\bibinfo {publisher} {Cambridge University Press,
  Cambridge},\ \bibinfo {year} {1998})%
  \bibAnnoteFile{NoStop}{Davies1998}%
\bibitem{APLWalther2006}%
  \BibitemOpen
  \bibfield{author}{%
  \bibinfo {author} {\bibfnamefont{C.}~\bibnamefont{Walther}}, \bibinfo
  {author} {\bibfnamefont{G.}~\bibnamefont{Scalari}}, \bibinfo {author}
  {\bibfnamefont{J.}~\bibnamefont{Faist}}, \bibinfo {author}
  {\bibfnamefont{H.}~\bibnamefont{Beere}},\ and\ \bibinfo {author}
  {\bibfnamefont{D.}~\bibnamefont{Ritchie}},\ }%
  \bibfield{journal}{%
  \Doi{10.1063/1.2404598}{\bibinfo {journal} {Appl. Phys. Lett.}}\ }%
  \textbf{\bibinfo {volume} {89}},\ \bibinfo {pages} {231121} (\bibinfo {year}
  {2006})%
  \bibAnnoteFile{NoStop}{APLWalther2006}%
\bibitem{Thesis_Valavanis2009}%
  \BibitemOpen
  \bibfield{author}{%
  \bibinfo {author} {\bibfnamefont{A.}~\bibnamefont{Valavanis}},\ }%
  \emph{\bibinfo {title} {{$n$-type silicon-germanium based terahertz quantum
  cascade lasers}}},\ Ph.D. thesis,\ \bibinfo {school} {School of Electronic
  and Electrical Engineering, University of Leeds} (\bibinfo {year} {2009}),\
  \url{http://etheses.whiterose.ac.uk/1262/}%
  \bibAnnoteFile{NoStop}{Thesis_Valavanis2009}%
\bibitem{PhysStatSol_Ko_2010}%
  \BibitemOpen
  \bibfield{author}{%
  \bibinfo {author} {\bibfnamefont{Y.~H.}\ \bibnamefont{Ko}}\ and\ \bibinfo
  {author} {\bibfnamefont{J.~S.}\ \bibnamefont{Yu}},\ }%
  \bibfield{journal}{%
  \Doi{10.1002/pssa.200925447}{\bibinfo {journal} {Phys. Stat. Sol. (a)}}\ }%
  \textbf{\bibinfo {volume} {207}},\ \bibinfo {pages} {2190} (\bibinfo {year}
  {2010})%
  \bibAnnoteFile{NoStop}{PhysStatSol_Ko_2010}%
\bibitem{Harrison2005}%
  \BibitemOpen
  \bibfield{author}{%
  \bibinfo {author} {\bibfnamefont{P.}~\bibnamefont{Harrison}},\ }%
  \emph{\bibinfo {title} {Quantum Wells, Wires and Dots}},\ \bibinfo {edition}
  {2nd}\ ed.\ (\bibinfo {publisher} {Wiley, Chichester},\ \bibinfo {year}
  {2005})%
  \bibAnnoteFile{NoStop}{Harrison2005}%
\bibitem{JOAValavanis2009}%
  \BibitemOpen
  \bibfield{author}{%
  \bibinfo {author} {\bibfnamefont{A.}~\bibnamefont{Valavanis}}, \bibinfo
  {author} {\bibfnamefont{Z.}~\bibnamefont{Ikoni\'c}},\ and\ \bibinfo {author}
  {\bibfnamefont{R.~W.}\ \bibnamefont{Kelsall}},\ }%
  \bibfield{journal}{%
  \Doi{10.1088/1464-4258/11/5/054012}{\bibinfo {journal} {J. Opt. A}}\ }%
  \textbf{\bibinfo {volume} {11}},\ \bibinfo {pages} {054012} (\bibinfo {year}
  {2009})%
  \bibAnnoteFile{NoStop}{JOAValavanis2009}%
\end{thebibliography}
%

\end{document}